\documentclass[twocolumn]{emulateapj}

\usepackage{color}
\usepackage{hyperref}

\usepackage{natbib}
\slugcomment{Accepted for publication in the Astrophysical Journal on December 28, 2015}

\shorttitle{Two Small Temperate Planets from $K2$}
\shortauthors{J. E. Schlieder et al.}

\begin{document}


\title{Two Small Temperate Planets Transiting Nearby M Dwarfs in $K2$ Campaigns 0 and 1\footnotemark[*,]\footnotemark[$\dagger$,]\footnotemark[$\ddagger$]}


\author{Joshua E. Schlieder\altaffilmark{1,2}, Ian J. M. Crossfield\altaffilmark{3,17}, Erik A. Petigura\altaffilmark{4,18}, Andrew W. Howard\altaffilmark{5}, Kimberly M. Aller\altaffilmark{5,2}, Evan Sinukoff\altaffilmark{5}, Howard  T. Isaacson\altaffilmark{6}, Benjamin J. Fulton\altaffilmark{5}, David R. Ciardi\altaffilmark{7}, Micka\"el Bonnefoy\altaffilmark{8}, Carl Ziegler\altaffilmark{9}, Timothy D. Morton\altaffilmark{10}, S\'ebastien L\'epine\altaffilmark{11}, Christian Obermeier\altaffilmark{12}, Michael C. Liu\altaffilmark{5}, Vanessa P. Bailey\altaffilmark{13}, Christoph Baranec\altaffilmark{14}, Charles A. Beichman\altaffilmark{7}, Denis Defr\`ere\altaffilmark{15}, Thomas Henning\altaffilmark{12}, Philip Hinz\altaffilmark{15}, Nicholas Law\altaffilmark{9}, Reed Riddle\altaffilmark{4}, Andrew Skemer\altaffilmark{15,16,18}}

\footnotetext[*]{Based in part on data obtained at the LBT. The LBT is an international collaboration among institutions in the United States, Italy and Germany. LBT Corporation partners are: The University of Arizona on behalf of the Arizona university system; Istituto Nazionale di Astrofisica, Italy; LBT Beteiligungsgesellschaft, Germany, representing the Max-Planck Society, the Astrophysical Institute Potsdam, and Heidelberg University; The Ohio State University, and The Research Corporation, on behalf of The University of Notre Dame, University of Minnesota and University of Virginia.}
\footnotetext[$\dagger$]{Some of the data presented herein were obtained at the W. M. Keck Observatory, which is operated as a scientific partnership among the California Institute of Technology, the University of California, and the National Aeronautics and Space Administration. The Observatory was made possible by the generous financial support of the W. M. Keck Foundation.} 
\footnotetext[$\ddagger$]{Based on observations collected at the European Organization for Astronomical Research in the Southern Hemisphere, La Silla Observatory, Chile during program ID 194.C-0443.}
\altaffiltext{1}{NASA Postdoctoral Program Fellow, NASA Ames Research Center, Space Science and Astrobiology Division, MS 245-6, Moffett Field, CA 94035, USA; \email{joshua.e.schlieder@nasa.gov}}
\altaffiltext{2}{Visiting Astronomer, NASA Infrared Telescope Facility}
\altaffiltext{3}{Lunar \& Planetary Laboratory, University of Arizona, 1629 E. University Blvd., Tucson, AZ, USA}
\altaffiltext{4}{California Institute of Technology, Pasadena, CA 91125, USA}
\altaffiltext{5}{Institute for Astronomy, University of Hawai`i, 2680 Woodlawn Drive, Honolulu, HI , USA}
\altaffiltext{6}{Astronomy Department, University of California, Berkeley, CA, USA}
\altaffiltext{7}{NASA Exoplanet Science Institute, California Institute of Technology, 770 S. Wilson Ave., Pasadena, CA 91125, USA}
\altaffiltext{8}{Universit\'e Grenoble Alpes, IPAG, 38000, Grenoble, 38000, Grenoble; CNRS, IPAG, 38000 Grenoble, France}
\altaffiltext{9}{University of North Carolina at Chapel Hill, Chapel Hill, NC 27599, USA}
\altaffiltext{10}{Department of Astrophysics, Princeton University, Princeton NJ, 08544, USA}
\altaffiltext{11}{Department of Physics \& Astronomy, Georgia State University, Atlanta, GA, USA}
\altaffiltext{12}{Max-Planck-Institut f\"ur Astronomie, K\"onigstuhl 17, 69117, Heidelberg, Germany}
\altaffiltext{13}{Kavli Institute for Particle Astrophysics and Cosmology, Stanford University, Stanford, CA 94305, USA}
\altaffiltext{14}{Institute for Astronomy, University of Hawai`i at M\={a}noa, Hilo, HI 96720-2700, USA}
\altaffiltext{15}{Steward Observatory, Department of Astronomy, University of Arizona, 933 N. Cherry Ave, Tucson, AZ 85721, USA}
\altaffiltext{16}{Department of Astronomy and Astrophysics, University of California, Santa Cruz, CA 95064, USA}
\altaffiltext{17}{Sagan Fellow}
\altaffiltext{18}{Hubble Fellow}






\begin{abstract}
{The prime $Kepler$ mission revealed that small planets ($<$ 4 R$_{\oplus}$) are common, especially around low-mass M dwarfs. $K2$, the re-purposed
 $Kepler$ mission, continues this exploration of small planets around small stars. Here we combine $K2$ photometry with spectroscopy, adaptive optics 
 imaging, and archival survey images to analyze two small planets orbiting the nearby, field age, M dwarfs K2-26 (EPIC 202083828) and K2-9. 
K2-26 is an $\mathrm{M1.0\pm0.5}$ dwarf at $93 \pm 7$ pc from $K2$ Campaign 0. We validate its 14.5665 d period planet and estimate a radius of  $\mathrm{2.67^{+0.46}_{-0.42}~R_{\oplus}}$. K2-9 is an $\mathrm{M2.5\pm0.5}$ dwarf at $110 \pm 12$ pc from $K2$ Campaign 1. K2-9b was first identified 
by \cite{MONTET15}; here we present spectra and adaptive optics imaging of the host star and independently validate and characterize the planet. Our analyses 
indicate K2-9b is a $\mathrm{2.25^{+0.53}_{-0.96}~R_{\oplus}}$ planet with a 18.4498 d period.  K2-26b exhibits a transit duration that is too long to be consistent with a circular orbit given the measured stellar radius. Thus, the long transits are likely due to the photoeccentric effect and our transit fits hint at  
an eccentric orbit.  Both planets receive low incident flux from their host stars and have estimated equilibrium temperatures $<$500 K. 
K2-9b may receive approximately Earth-like insolation. However, its host star exhibits strong $GALEX$ UV emission which could affect any atmosphere it harbors.    
K2-26b and K2-9b are representatives of a poorly studied class of small planets with cool 
temperatures that have radii intermediate to Earth and Neptune. Future study of these systems can provide key insight into trends in 
bulk composition and atmospheric properties at the transition from silicate dominated to volatile rich bodies.}
\end{abstract}





\section{Introduction}

Planets are commonplace in the Galaxy. In the last 20 years, knowledge of planet demographics, architectures, and frequencies
has expanded beyond the eight primary bodies in our solar system to thousands of planets orbiting thousands of stars. A
workhorse of this exoplanet revolution is the $Kepler$ space telescope. Transit data collected during the prime mission of $Kepler$
revealed that small planets: Earth analogues, super-Earths, and sub-Neptunes  (R$_p$ $<$ 4 R$_{\oplus}$), are abundant 
around Sun-like stars \citep{PETIGURA13_PNAS}. Statistical studies
focusing on the few M dwarfs 
($\mathrm{T_{eff} \lesssim 4000~K,~M_* \lesssim 0.6~M_{\odot}}$) that  
$Kepler$ observed (3900 stars) revealed that small planets exist around \emph{nearly all} M dwarfs \citep{DRESSING13, DRESSING15}.  

The small radii and masses of M dwarfs, combined with their sheer numbers \citep[$\sim$70\% of all stars,][]{BOCHANSKI10}, provide the best 
opportunities to detect and characterize small planets in the Solar neighborhood. Because of the large numbers of M dwarfs and the high frequency of 
planets around them, the closest Earth-size planets in the habitable zone almost certainly orbit these low-mass stars. Planets of a given radius transiting M 
dwarfs exhibit deeper transit signatures and planets of a given mass produce larger stellar reflex motions \citep{HOWARD12}. 
Additionally, the atmospheres of small planets orbiting M dwarfs are more amenable to transmission spectroscopy studies \citep[e.g.][]{KREIDBERG14} 
due to the favorable star-to-planet radius ratio. However, since $Kepler$ observed relatively few of these stars, the number of small planets detected and confirmed 
in transit around M dwarfs remains small. Subsequently, their demographics, formation scenarios, and the evolution of their orbits remain poorly constrained.

We are pursuing a program to identify and characterize additional small planets transiting M dwarfs using data from $K2$, the 
2 reaction wheel, ecliptic plane survey of NASA's re-purposed \emph{Kepler} spacecraft \citep{HOWELL14}. The $K2$ M Dwarf Program
($K2$-MDP) is a comprehensive approach to select M dwarf targets in each $K2$ field, generate calibrated light curves 
and identify candidate transiting planets, and obtain follow-up observations to validate and characterize 
the planetary systems. The first discoveries from the $K2$-MDP are K2-3 and K2-21, M0 dwarfs within 100 pc each hosting multiple transiting 
super-Earths \citep{CROSSFIELD15, PETIGURA15}.  K2-3bcd, K2-21bc, and other early $K2$ discoveries \citep[][]{VANDERBURG15, MONTET15} 
have provided planets that occupy poorly explored regions of the planetary 
mass-radius-temperature diagram ($\mathrm{R_p < 4 R_{\oplus}, T_{eq} < 600 K}$), some ideal early targets for spectroscopic follow-up with the 
\emph{James Webb Space Telescope} \citep[\emph{JWST},][]{BATALHA15, BEICHMAN14}, and some truly novel systems \citep[i.e. WASP-47bcd,][]{BECKER15}; 
all well before the launch of the \emph{Transiting Exoplanet Survey Satellite} \citep[\emph{TESS},][]{RICKER14}.

Here we present the discovery and validation of a small, cool planet orbiting the nearby M dwarf K2-26 
and an independent validation and detailed characterization of the known planet transiting the M dwarf K2-9. In \S 2 we describe the observations 
of these systems using $K2$ and ground based spectroscopy and imaging. We detail our analyses of these observed 
data and the results in \S 3. \S 4 provides a discussion of the properties of these planets in the context of known demographics 
and \S 5 provides concluding remarks.

\setcounter{footnote}{0}

\begin{table*}[!t]
\begin{center}
\caption{Summary of Stellar Properties \label{tab1}}
\begin{tabular}{lccc}
\tableline\tableline
Parameter & K2-26 & K2-9 & Reference \\
\tableline
$\alpha$ (hh:mm:ss) & 06:16:49.579   & 11:45:03.472  & 1 \\ 
$\delta$ (dd:mm:ss) & +24:35:47.08 &  +00:00:19.08  & 1 \\
$\mu_{\alpha}$ (mas yr$^{-1}$) & $-27.8\pm4.1$  & $-171.6\pm3.8$   & 2 \\
$\mu_{\delta}$ (mas yr$^{-1}$) & $-117.9\pm4.1$  &  $32.1\pm3.8$   & 2 \\
RV (km s$^{-1}$) & $95.34\pm0.15$ & -$31.02\pm0.15$ & 1 \\
d$_{phot}$ (pc) &  $93\pm7$ & $110\pm12$ &  1 \\
\tableline
$Kep$ (mag) & 14.00  & 14.96  &  1  \\
$B$(mag) &  $15.97\pm0.13$  &  \dots  &  3  \\
$V$(mag) &  $14.53\pm0.03$  &  15.63$^a$  &  3  \\
$B_{POSSI}$ (mag) &  16.16  &  16.55  &  2  \\
$R_{POSSI}$ (mag) &  13.14  &   14.41  &  2  \\
$g^{\prime}$ (mag) &  $15.296\pm0.023$  & $16.652\pm0.117$  &  3\\
$r^{\prime}$ (mag)&  $13.927\pm0.080$  &  $15.218\pm0.018$  &  3\\
$i^{\prime}$ (mag) &  $13.421\pm0.493$  &  $14.147\pm0.095$ &  3  \\
$J$ (mag) & $11.350\pm0.024$  & $12.451\pm0.024$ & 4 \\
$H$ (mag) & $10.762\pm0.022$  & $11.710\pm0.022$ & 4 \\
$K_{s}$ (mag) &  $10.530\pm0.018$  &  $11.495\pm0.023$   &  4 \\
$W1$ (mag) & $10.422\pm0.023$  & $11.348\pm0.022$ & 5 \\
$W2$ (mag) & $10.349\pm0.021$  & $11.214\pm0.021$ & 5 \\
$W3$ (mag) & $10.409\pm0.086$  & $11.354\pm0.193$ & 5 \\
\tableline
Spectral Type   &  M1.0V $\pm$ 0.5   &   M2.5V$\pm$0.5  & 1 \\
$\mathrm{T_{eff}}$ (K)   &   $3785\pm185$   &   $3390\pm150$ & 1 \\
$\mathrm{[Fe/H]}$ (dex) &  $-0.13\pm0.15$ & $-0.25\pm0.20$ & 1 \\
Radius ($\mathrm{R_{\odot}}$)  &  $\mathrm{0.52\pm0.08}$ &  $\mathrm{0.31\pm0.11}$   &    1  \\
Mass ($\mathrm{M_{\odot}}$)  &  $\mathrm{0.56\pm0.10}$ &  $\mathrm{0.30\pm0.14}$   &    1  \\
Luminosity ($\mathrm{L_{\odot}}$) &  $\mathrm{0.049\pm0.023}$ &  $\mathrm{0.012\pm0.010}$   &    1  \\
Density (g cm$^{-3}$) &  $\mathrm{3.92\pm1.43}$ &  $\mathrm{9.88\pm4.25}$   &    1  \\
Age (Gyr) & $\gtrsim$1 & $\gtrsim$1  & 1 \\ 
\tableline
\label{properties_table}
\end{tabular}
\tablecomments{1 - This Work; 2 - \citet[][PPMXL]{ROESER10}; 3 - \citet[][APASS DR9]{HENDEN12}; 4 - \citet[][2MASS]{CUTRI03};  5 - \citet[][ALLWISE]{CUTRI13}; $^a$ - Estimated using photometric relations in \cite{LEPINE05}}
\end{center} 
\end{table*}
 
\section{Observations and Data Reduction}

\subsection{$K2$ Target Selection, Photometry, and Transit Search}

We identified the high proper motion stars PM I06168+2435 and PM I11450+0000 
(LP 613-39, NLTT 28423) as candidate M dwarf targets for our $K2$ Campaign 0 (C0: GO0120 - PI L\'epine) 
and Campaign 1 (C1: GO1036 - PI Crossfield) proposals, respectively. {\color{black} The stars were also proposed as targets in 
C0 and C1 by several other groups (C0: GO0111 - PI Sanchis Ojeda, GO0119 - PI Montet; C1: GO1052 - PI Robertson, GO1053 - PI Montet, 
GO1059 - PI Stello, GO1062 - PI Anglada-Escude). We selected these targets as} candidate nearby M dwarfs
from the SUPERBLINK proper motion survey \citep[][]{LEPINE05, LEPINE11} following
the photometric and proper motion criteria described in \cite{CROSSFIELD15}.  
A coordinate cross-match of PM I06168+2435 and PM I11450+0000 with the $K2$ 
\emph{Ecliptic Plane Input Catalog} (EPIC) returned matches with the sources
EPIC 202083828 and EPIC 201465501, respectively.  EPIC 201465501 was given the $K2$ identifier
K2-9 by NExScI\footnote{\url{http://exoplanetarchive.ipac.caltech.edu/docs/K2Numbers.html}} 
after validation of its planet in \cite{MONTET15}. EPIC 202083828 was designated K2-26 after the validation of its 
planet in \S~3.4 of this work.  K2-26 was observed by $K2$ in long-cadence mode during
C0 from 2014 March 08 to May 27 and K2-9 was observed using the same mode
during C1 from 2014 May 30 to August 21. We provide basic identifying information and 
available photometry for these stars in Table~\ref{properties_table}. 

\begin{figure*}[!htb]
\epsscale{1.2}
\plotone{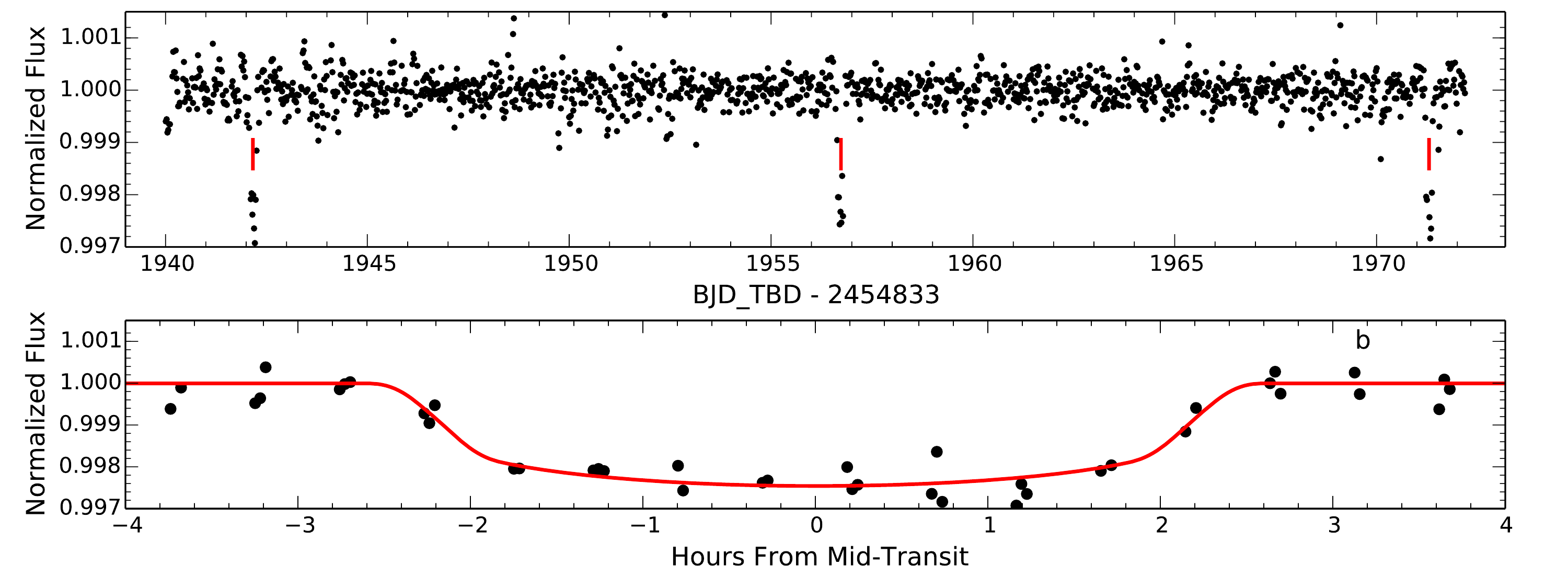}
\caption{$Top$: Calibrated $K2$ photometry for K2-26 (EPIC 202083828). Vertical ticks indicate the locations of the transits. $Bottom$: Phase-folded photometry
and best-fit light curve. The $\sim$4.7 h transit duration is likely the result of an eccentric orbit. \label{3828_lightcurve_fig}}
\end{figure*}

\begin{figure*}[!htb]
\epsscale{1.2}
\plotone{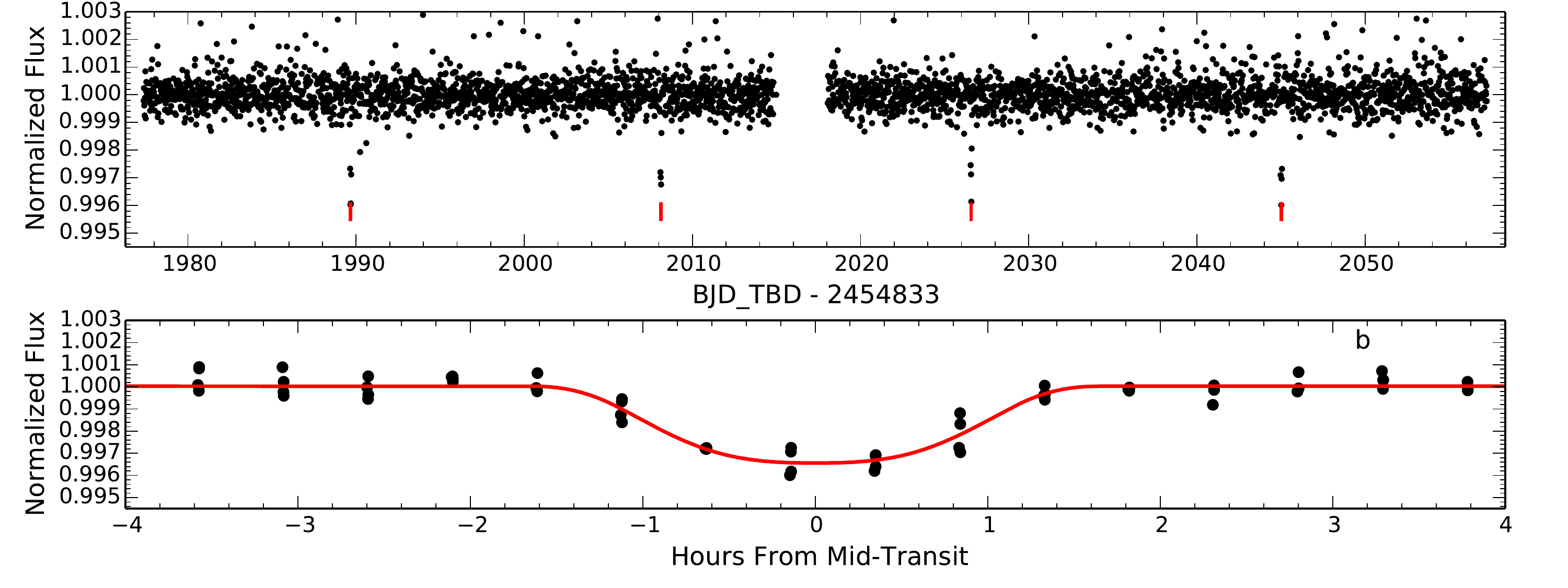}
\caption{$Top$: Calibrated $K2$ photometry for K2-9. Vertical ticks indicate the locations of the transits. $Bottom$: Phase-folded photometry
and best-fit light curve.   \label{5501_lightcurve_fig}}
\end{figure*}

The degraded pointing precision of $K2$ due to the loss of 2 reaction wheels leads to telescope drift in 
the form of a roll around the telescope boresight. This drift is corrected using thruster fires when 
the space craft reaches a predetermined roll limit; approximately every 6 h.  The periodic drift and correction of a star over 
$\sim$1 pixel leads to systematic brightness variations of $\sim$0.5\%. These variations are roll angle dependent
and must be corrected in the light curve extraction process. Our approach to correcting these effects 
and extracting calibrated photometry from the raw $K2$ pixel data is identical to that described in 
\cite{CROSSFIELD15}. In general, we perform a frame by frame median flux subtraction and compute 
the the raw photometry by summing the single frame flux within a circular aperture centered on the target.
We then compute the principal components of the row and column centroids and
fit a Gaussian process (GP) to remove the systematic variations. In practice, the GP fitting is iterative 
and updates are made to the GP parameters to minimize the rms fit residuals. The flux extraction and GP fitting procedures
are repeated for apertures of varying size until an aperture size is found that minimizes the rms residuals in the 
calibrated light curve. The extraction apertures for K2-26 and K2-9 were soft-edged, circular apertures having radii of 2 and 3 pixels, respectively. 
After correcting for spacecraft roll, both stars exhibit smooth, low-amplitude, slowly modulating, photometric variations on the order of 1\%. These features could be 
related to intrinsic stellar variability or unaccounted for spacecraft systematics; possibly small focus changes due to 
thermal expansion and contraction over the course of a $K2$ observing campaign. If the observed residual variability is at least 
in part intrinsic to the stars, i.e. star spots, it is very low-level and indicative of slow rotation rates in both cases ($\sim$weeks). Prior to searching for transit 
events in the light curves, this residual variability is also removed. The calibrated $K2$ light curves for the stars are shown in top panels of 
Figures~\ref{3828_lightcurve_fig} and~\ref{5501_lightcurve_fig} and are available upon request.

We searched the calibrated and detrended light curves of K2-26 and K2-9 using the \texttt{TERRA} algorithm; an automated, grid-based, transit search pipeline
 \citep{PETIGURA12, PETIGURA13a}. Our \texttt{TERRA}
search of the K2-26 photometry identified a candidate planet with a period of $P\approx14.567$ d and SNR~$\approx$~34. A candidate was also detected transiting
K2-9 with $P\approx18.450$ d and SNR~$\approx$~24. Each of these transit signals was fit with a \citet{MANDEL02} transit model 
which we show in the bottom panels of Figures~\ref{3828_lightcurve_fig} and~\ref{5501_lightcurve_fig}. We then masked out the in-transit observations of each planet candidate
and searched for additional transit signals with \texttt{TERRA}. This subsequent search of each light curve returned no further candidates above our 
SNR threshold of 12.  {\color{black} We note that the maximum likelihood periods for the candidates transiting K2-26 and K2-9 are close to 
integer multiples of the observing cadence. We consider this a priori unlikely, and hypothesize that residual systematics exist in the \emph{K2} photometry after 
our photometric processing. The \emph{K2} C0-C3 candidate catalog of \cite{VANDERBURG_15_cat} includes K2-26 and K2-9 with periods of 14.5670 d 
and 18.4487 d (no uncertainties), respectively. The \cite{VANDERBURG_15_cat} photometry was extracted using an independent analysis and they find periods 
consistent with ours at the $\sim$0.3$\sigma$ and $\sim$0.8$\sigma$ levels. \cite{MONTET15} also report an independent period for K2-9 of $18.44883\pm0.00137$ d, 
consistent with our estimate at the $\sim$1$\sigma$ level. Thus, we conclude that any systematic errors that favor periods that are near-integer multiples of the $K2$ 
long cadence are second order and have minimal impact on our reported parameters.} To validate and characterize these candidate planets, we obtain and analyze 
spectroscopic and imaging data, perform detailed checks of the $K2$ pixel level photometry and light curves, and estimate false positive probabilities. These observations 
and analyses are described in the following sections.

\subsection{Follow-up Spectroscopy}

\subsubsection{IRTF/SpeX}

We observed K2-26 and K2-9 using the near-infrared cross-dispersed 
spectrograph \citep[SpeX, ][]{RAYNER03} on the 3.0m NASA Infrared Telescope facility on 2015 
May 02 UT and 2015 April 16 UT, respectively. K2-26 was observed under clear
skies with an average seeing of $\sim$0\farcs6. K2-9 was observed under poorer conditions 
with thin, variable cirrus, high humidity, and seeing between 1\farcs0 - 1\farcs2. We used the instrument 
in short cross dispersed mode using the $0.3 \times 15^{\prime\prime}$ slit which provides wavelength 
coverage from 0.68 to 2.5 $\mu$m at a resolution of $R \approx 2000$. The stars were dithered to two 
positions along the slit following an AB pattern for sky subtraction. The K2-26 observing sequence 
consisted of $8 \times 75$ s exposures for a total integration time of 600 s. K2-9 was observed for
$24 \times 120$ s for a total time of 2880 s. We also observed an A0 standard and flat and arc lamp 
exposures immediately after each star for telluric correction and wavelength calibration.

\begin{figure*}[!htb]
\epsscale{1.2}
\plotone{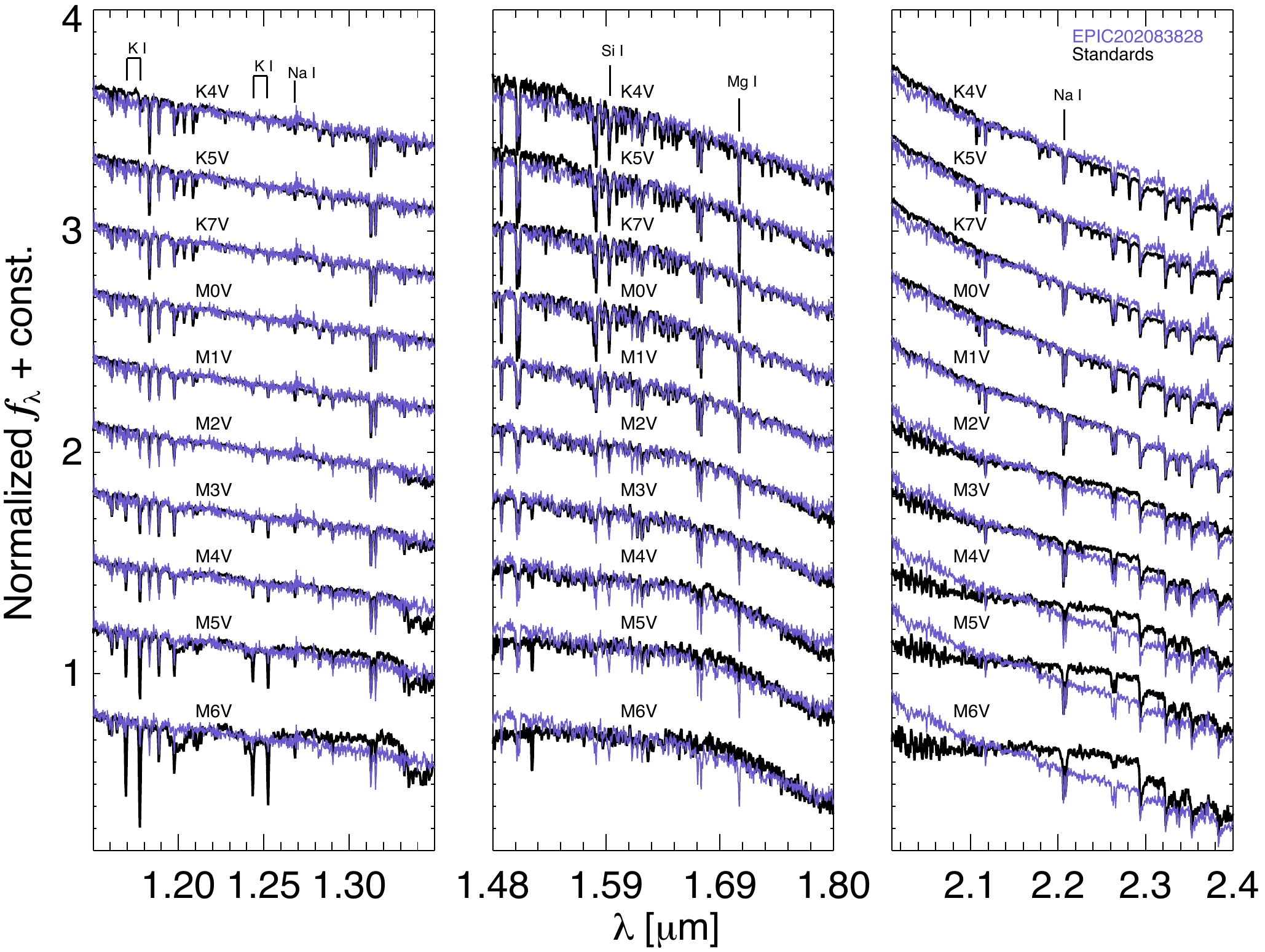}
\caption{$JHK$-band IRTF/SpeX spectra of K2-26 (EPIC 202083828) compared to late-type standards from the IRTF spectral library. 
All spectra are normalized to the continuum in each of the plotted regions. The star is a best visual match to spectral type M1 across 
the three near-IR bands. This is consistent with the results from our analyses using spectroscopic indices.  \label{SpeX_fig1}}
\end{figure*}

\begin{figure*}[!htb]
\epsscale{1.2}
\plotone{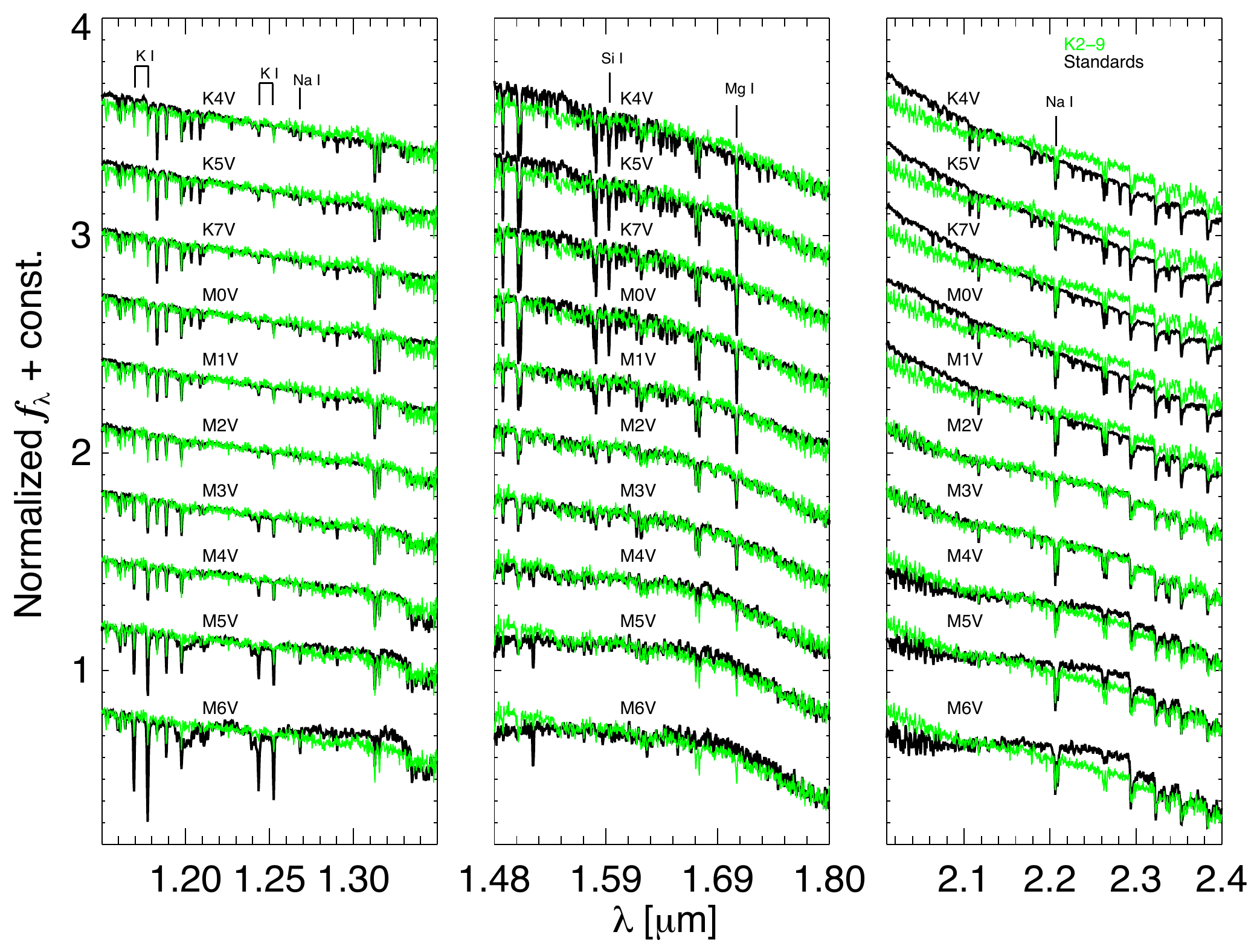}
\caption{$JHK$-band IRTF/SpeX spectra of K2-9 compared to late-type standards from the IRTF spectral library. All spectra 
are normalized to the continuum in each of the plotted regions. The star is a best visual match to spectral type M2/M3 across the three near-IR bands. 
This is consistent with the spectral type derived from spectroscopic index based methods.  \label{SpeX_fig2}}
\end{figure*}

The data were reduced using the SpeXTool package \citep{VACCA03, CUSHING04}.  SpeXTool performs 
flat fielding, bad pixel removal, wavelength calibration, sky subtraction, spectral extraction and combination, 
telluric correction, flux calibration, and order merging. The final calibrated spectra had signal-to-noise ratios 
(SNR) of $\sim$80 per resolution element in the $H$- ($\sim$1.6 $\mu$m) and $K$-bands ($\sim$2.2 $\mu$m). The spectral
quality decreases rapidly toward bluer wavelengths with SNR $\sim$60 in the $J$-band ($\sim$1.25 $\mu$m) and $\sim$10 
at 0.75 $\mu$m.  The $JHK$-band spectra are compared to late-type standards from the IRTF 
Spectral Library\footnote{\url{http://irtfweb.ifa.hawaii.edu/~spex/IRTF\_Spectral_Library/}}
\citep{CUSHING05, RAYNER09} in Figures~\ref{SpeX_fig1}~and~\ref{SpeX_fig2}. K2-26 is a best visual 
match to the M1 standard across the near-IR bands. K2-9 is later-type and matches well with the M2/M3 
standards.

\subsubsection{NTT/EFOSC2}

On UT 2015 January 11, we observed K2-26 using the ESO Faint Object Spectrograph and Camera (v.2) 
\citep[EFOSC2,][]{BUZZONI84} mounted to the Nasmyth B focus of the 3.6m ESO New Technology Telescope 
(NTT). These observations were made as part of our 70 night $K2$ follow-up program (PID 194.C-0443, PI: I.J.M. Crossfield). 
The star was observed under good conditions with average seeing $\sim$1\farcs0 with a total integration time
of 270 s. We used EFOSC2 in spectroscopic mode with the 0\farcs3 slit and grism 16 to provide a 
resolution $R\sim1600$ from 0.6-1.0 $\mu$m. We also obtained standard bias, flat, and HeAr lamp calibration 
frames immediately after observing K2-26 along with observations of spectrophotometric standards for 
flux calibration \citep{BOHLIN01}.

The EFOSC2 data was reduced using standard IRAF\footnote{IRAF is distributed by the National Optical 
Astronomy Observatories, which are operated by the Association of Universities for Research in Astronomy, 
Inc., under cooperative agreement with the National Science Foundation.} routines that included bias subtraction, 
flat fielding, wavelength calibration, and spectral extraction. The spectrum was then flux calibrated using a 
standard observed close in time. The final calibrated spectrum had a SNR $\sim$50 per resolution element. 

\subsubsection{Palomar Hale 5.0m/Double Spectrograph}       

\begin{figure*}
\epsscale{1.15}\plottwo{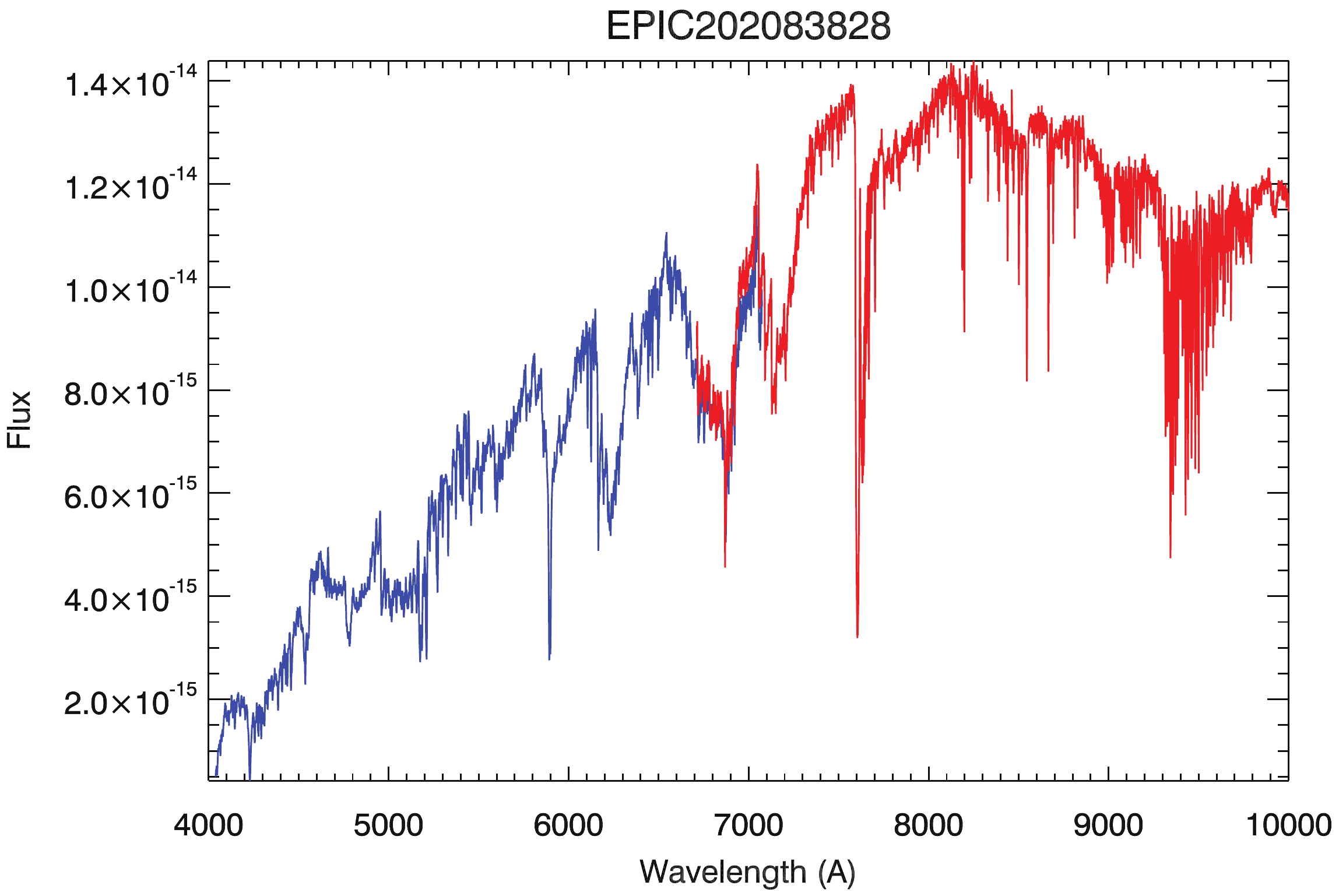}{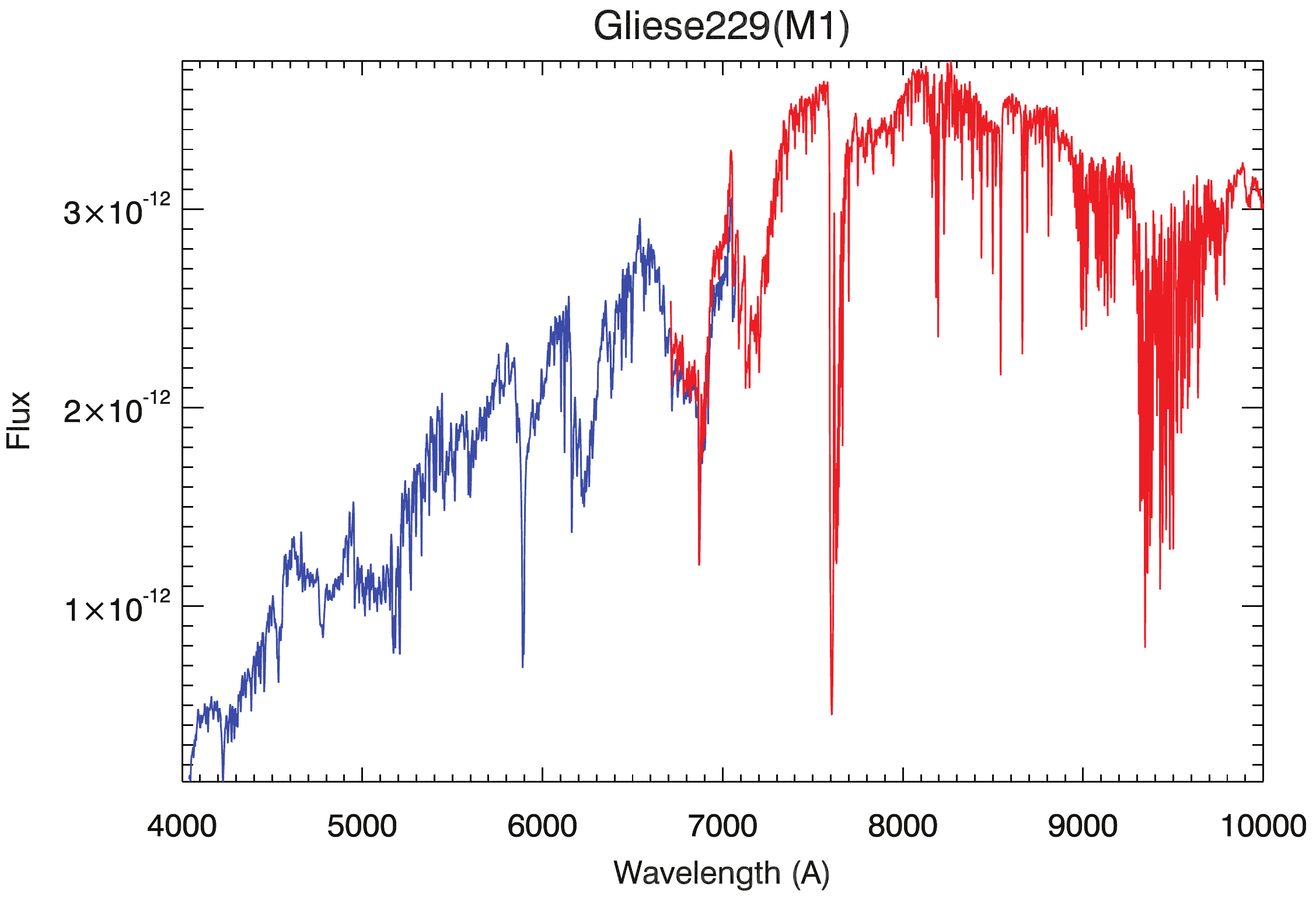}
\caption{$Left$: Spectra of K2-26 (EPIC 202083828) taken with the blue and red sides of the Double Spectrograph (DBSP) at the Palomar Hale 5.0 m. The flux units are arbitrary. $Right$: A comparison spectrum of the M1 standard star GJ 229. The overall continuum shape and strength of the deep, broad molecular features (TiO, CaH, VO) of K2-26 are an excellent match to the M1 standard across the observed wavelength range.  \label{DBSP}}
\end{figure*}

We observed K2-26 using the Double Spectrograph \citep[DBSP,][]{OKE82} at the Palomar observatory Hale 5.0m telescope on 2015 February 12 UT. 
On the blue side of the spectrogaph, the 600 l/mm 
grating blazed at 3780~\AA~was used at a setting of 29.5$^{\circ}$. On the red side, the 600 l/mm grating blazed at 9500~\AA~was used at an angle of 32.5$^{\circ}$. 
The star was observed with a 1\arcsec\ slit which provided a spectral resolution of R$\sim$2400 and wavelength coverage from $\sim$4000 - 7000~\AA~on 
the blue side and R$\sim$3000 and coverage from $\sim$7000 - 10000~\AA~on the red side. The target was 
observed at an airmass of 1.0 and conditions were generally favorable with seeing of about 1.5-1.8\arcsec. Standard IRAF functions ($apall$,  $standard$, $sensfunc$)\footnote{\url{http://www.twilightlandscapes.com/IRAFtutorial/IRAFintro\_06.html}} were used to calibrate the data including: bias frame subtraction and flat-fielding using 
dome flats,  wavelength calibration using Fe-Ar arcs in the blue and He-Ne-Ar arcs in the red, and initial flux calibration with respect to standard Hiltner 600 \citep{HAMUY94}.  
A separate IDL routine was used to stitch together the red and blue spectra. A 4000-10000~\AA~portion is shown in Figure~\ref{DBSP}. For comparison, we also 
show the spectrum of the M1 standard star GJ 229 \citep{KIRKPATRICK91, MALDONADO15} observed using the same DBSP settings. 

\subsubsection{Keck/HIRES}

We observed both stars using the High Resolution Echelle Spectrometer \citep[HIRES,][]{VOGT94} on the 10.0m Keck I telescope. We observed the stars following 
standard California Planet Search \citep[CPS,][]{MARCY08} procedures using the C2 decker and the 0\farcs87~$\times$~14\farcs0 slit. The
0\farcs87 slit provides wavelength coverage from $\sim$3600 - 8000~\AA~at a resolution $R \approx 60000$. No Iodine cell was used for these observations, the wavelength
scale was calibrated using the standard HIRES reference. {\color{black} K2-26 was observed on UT 2015 February 5 and UT 2015 November 15 under good conditions with $\sim$1\farcs0 seeing for a total of 565 s on each night.}
K2-9 was observed on UT 2015 July 12 under clear skies with 1\farcs4 seeing for a total of 1200 s. These data were reduced using the standard pipeline of the 
CPS \citep[][]{MARCY08}. The resulting spectra of K2-26 and K2-9 had SNR's~$\sim$30 
and 25 per pixel at 5500~\AA, respectively. Examples of the HIRES spectra for both stars are shown in Figure~\ref{HIRES_fig}.

\subsection{Adaptive Optics and Archival Imaging}

\subsubsection{LBT - LBTI/LMIRcam}

\begin{figure*}[!htb]
\epsscale{1.2}
\plotone{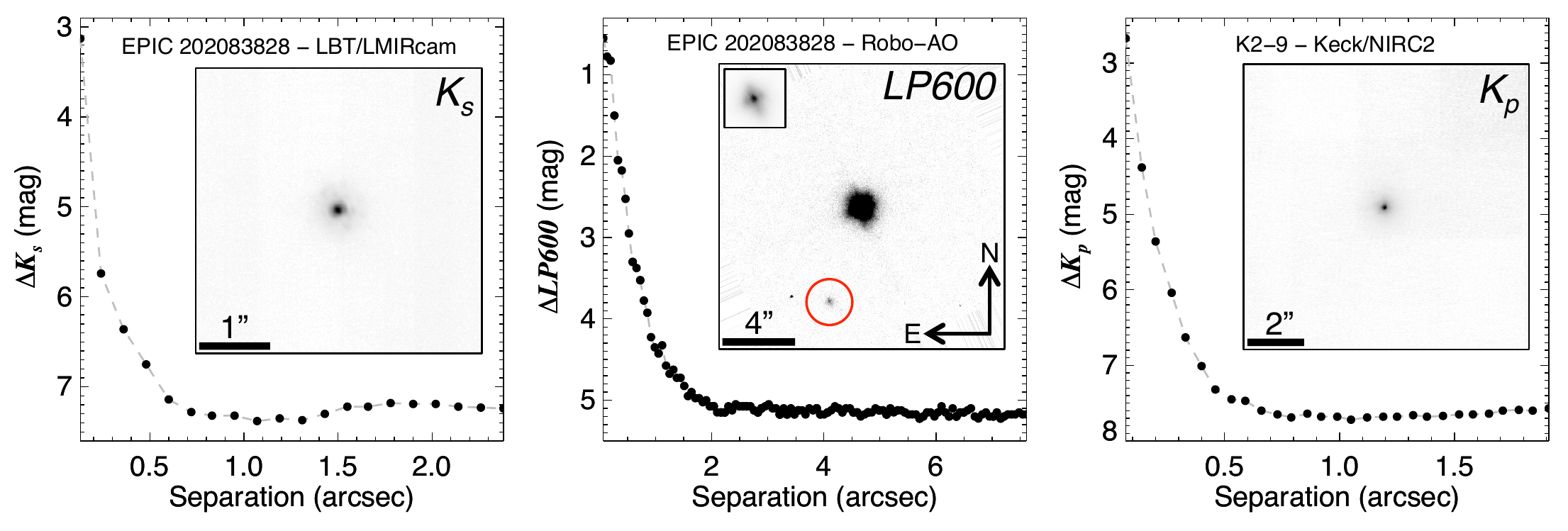}
\caption{AO images and contrast curves for K2-26 (EPIC 202083828) and K2-9 $Left$: K2-26 LBT/LMIRcam $K_s$-band image (inset) and contrast curve. 
No additional stars are detected within 2\farcs0.
$Center$: K2-26 Robo-AO $LP600$-band image (inset) and contrast curve. An additional star is detected with 
$\Delta LP600 = 5$ mag at 5.5$^{\prime\prime}$ separation (red circle). The small box in the upper left of the inset shows K2-26 
without the hard stretch necessary to reveal the faint companion. $Right$: K2-9 Keck/NIRC2
$K_p$-band image and contrast curve. No additional stars are detected in the NIRC2 field of view. \label{AO_fig}}
\end{figure*}

K2-26 was observed on 2015 January 07 UT using the $L/M$-band Infrared Camera  
\citep[LMIRcam,][]{SKRUTSKIE10, LEISENRING12} of the LBT Interferometer \citep[LBTI,][]{HINZ08}. 
LBTI/LMIRcam is mounted at the bent Gregorian focus of the dual 8.4m Large Binocular Telescope (LBT) and works in 
conjunction with the deformable secondary LBT Adaptive Optics system \citep[LBTIAO,][]{ESPOSITO10, ESPOSITO11, RICCARDI10, BAILEY14} 
to deliver high-resolution near-IR imaging. For our observations, we only used the right side of the LBT. K2-26 
was observed using the $K_s$-band filter ($\lambda_c=2.16$ $\mu$m, $\Delta\lambda=0.32$ $\mu$m) following a two 
point dither pattern for sky subtraction. We obtained $40\times0.15$s exposures using the target as a natural 
AO guide star for a total integration time of 6s. Our data reduction included corrections for detector bias, sky background, and bad pixels followed by
frame re-centering and averaging. The reduced image has a field-of-view (FOV) 10\farcs9 and a plate scale
of $10.707\pm0.012$ mas pixel$^{-1}$ \citep{MAIRE15}. To ensure optimal background subtraction and 
contrast, the final image of K2-26 is trimmed to a 4\farcs0 region of full dither overlap. This is 
shown in the inset of the left panel of Figure~\ref{AO_fig}.

\subsubsection{Palomar 60 Inch/Robo-AO}

We acquired visible-light adaptive optics images of K2-26 using the Robo-AO system \citep{BARANEC13, BARANEC14} on the 60-inch Telescope 
at Palomar Observatory. On 2015 March 8 UT, we observed K2-26 with a long-pass filter cutting on at 600 nm ($LP600$) as a sequence of 
full-frame-transfer detector readouts from an electron-multiplying CCD at the maximum rate of 8.6 Hz for a total of 120 s of integration time. 
The individual images are corrected for detector bias and flat-fielding effects before being combined using post-facto shift-and-add processing 
using K2-26 as the tip-tilt star with 100\% frame selection to synthesize a long-exposure image \citep{LAW14}. The resulting reduced image 
has a nominal FOV of 44\farcs0 and plate scale of 0\farcs0216 pixel$^{-1}$ \citep{BARANEC14}. A 15\farcs5 portion of the Robo-AO image 
centered on K2-26 is shown in the inset of the center panel of Figure~\ref{AO_fig}. A faint, widely separated companion was detected
in the Robo-AO image and is described in \S~\ref{AO_sec}.

\subsubsection{Keck/NIRC2}

We observed K2-9 using the Near Infrared Camera 2 (NIRC2) and laser guide star AO \citep[LGS AO,][]{VANDAM06, WIZINOWICH06} 
 on the 10.0m Keck-II telescope on 2015 April 07 UT.  
The target was observed in the $K_p$-band filter  ($\lambda_c=2.124$ $\mu$m, 
$\Delta\lambda=0.351$ $\mu$m) using the narrow camera setting with a pixel scale of 9.942 mas pixel$^{-1}$. To avoid the noisier lower-left 
quadrant of the NIRC2 array, we employed a three-point dither 
pattern with $11\times10$s integrations per dither yielding a total on-source integration time
of 330s. Individual frames were flat-fielded and sky-subtracted and then shifted and coadded to produce the final 10\farcs2 image.
Our NIRC2 image of K2-9 is shown in the inset of the right panel of Fig.~\ref{AO_fig}.

\subsubsection{DSS and SDSS Archival Imaging}

\begin{figure*}[!htb]
\epsscale{1.1}
\plotone{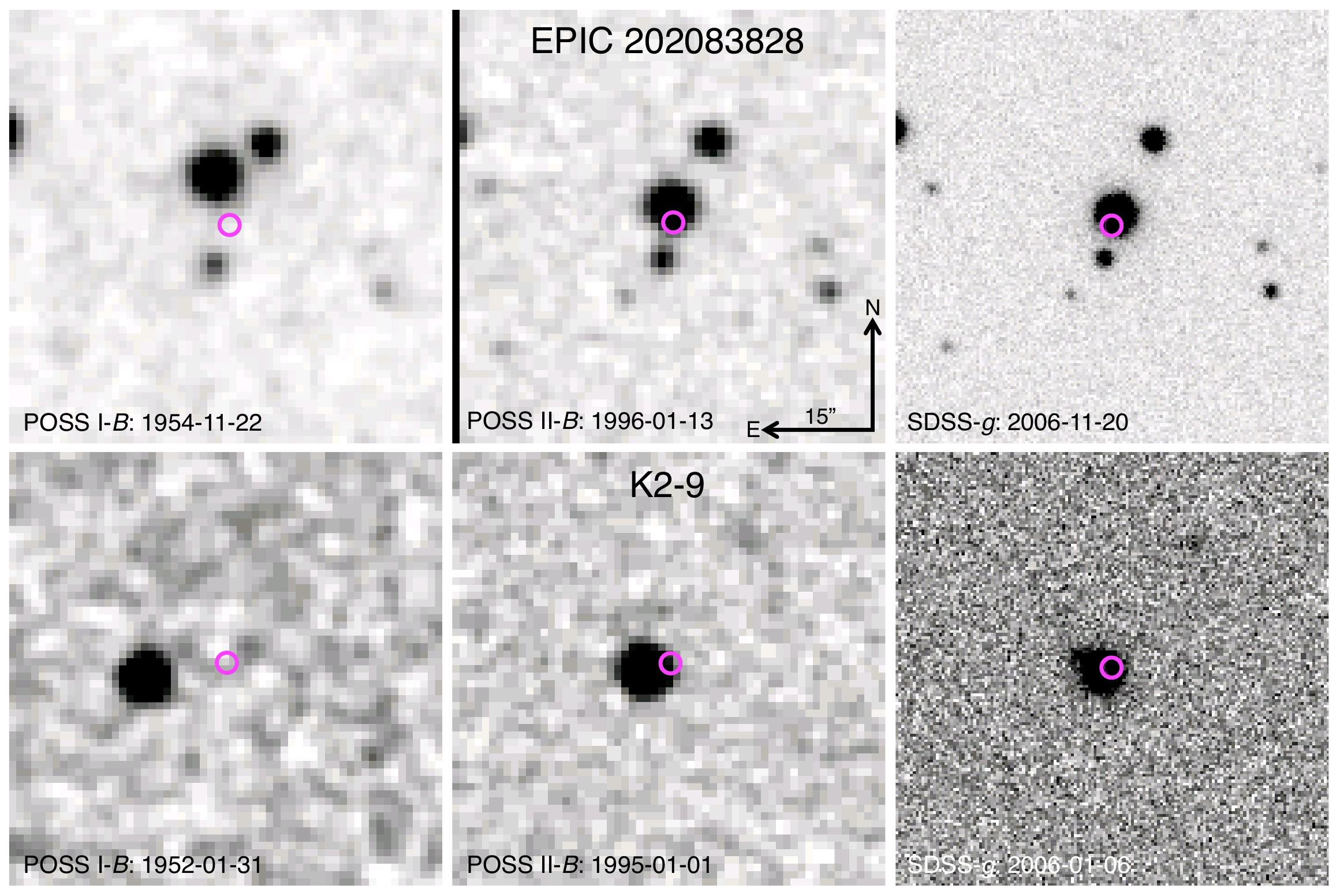}
\caption{1\farcm0 $\times$ 1\farcm0 archival survey images of K2-26 (EPIC 202083828) and K2-9. $Top$: K2-26 DSS and SDSS images displaying 
$>$6$^{\prime\prime}$ of transverse motion over 52 years. The faint star to the 
south-southeast in each image is the same as revealed in our Robo-AO images. The star is not co-moving with K2-26. $Bottom$: DSS and SDSS 
images of K2-9 showing $>$9$^{\prime\prime}$ of transverse motion over 54 years. A very faint source is detected  $\sim$3\farcs5 to the northeast of  K2-9 in the SDSS image and appears 
as an extension of the K2-9 intensity distribution. This source is not detected with confidence in the shallower POSS images. Further details are provided in section~\ref{archival_sec}. The POSS I images reveal no stars at the current positions of K2-26 or K2-9  (magenta circles) down to the photometric limits of that survey. \label{archival_fig}}
\end{figure*}

K2-26 and K2-9 were both observed in two different photometric bands (blue and red; $B$ and $R$) during 
The National Geographic Society - Palomar Observatory Sky Survey \citep[POSS I,][]{MINKOWSKI63} and the 
Second Palomar Observatory Sky Survey \citep[POSS II,][]{REID91} using the 1.2m Samuel Oschin Telescope. The original
POSS photographic plates were scanned and digitized by the Space Telescope Science Institute and are now available for flexible 
download as the Digitized Sky Survey (DSS)\footnote{\url{http://stdatu.stsci.edu/cgi-bin/dss\_form}}.
The digitized POSS I and II plates have plate scales of 1\farcs01 pixel$^{-1}$. {\color{black} K2-26 was observed in the $B$ and $R$-bands during 
POSS I on 1954 November 22 UT and in the $B$-band during POSS II on 1996 January 13 UT. K2-9
was observed in both POSS I bands on 1952 January 31 UT and in the POSS II $B$-band on 1995 January 01 UT.} Both stars were also observed in five photometric bands 
($u, g, r, i, z$) during the Sloan Digital Sky Survey \citep[SDSS,][]{YORK00} using the
2.5m Sloan Foundation Telescope \citep{GUNN06}. SDSS images have a plate scale of 0\farcs396 pixel$^{-1}$. K2-26 and K2-9 
were observed during the SDSS on 2006 November 11 UT
and 2006 January 06 UT, respectively. The total time baselines between the POSS I and SDSS epochs for each star are 52 and 54 years, respectively. 
We obtained the publicly available imaging data in the form of 1\farcm0 POSS I $B$, POSS II $B$, and SDSS DR7 \citep{ABAZAJIAN09} $g$ images of the stars 
centered on their 2015 epoch positions using the NASA/IPAC Infrared Science Archive (IRSA) Finder Chart web 
interface\footnote{\url{http://irsa.ipac.caltech.edu/applications/finderchart/}}. These data are presented in Figure~\ref{archival_fig}.

\section{Analyses and Results}

\subsection{Spectroscopic Analyses}

\subsubsection{Medium-Resolution Spectroscopy}

We use molecular band indices in the optical and near-IR to estimate spectral types (SpTy) for K2-26 and K2-9. The EFOSC2 spectrum
of K2-26 provides access to TiO and CaH molecular bands that are temperature sensitive and calibrated to provide SpTy's for stars $\sim$K7-M6. We specifically
use the TiO5, CaH2, and CaH3 indices \citep{REID95, GIZIS97} to estimate the star's SpTy using the calibrated relations in \citet{LEPINE13}. We find K2-26 has an 
optical SpTy of M1.0. The \citet{LEPINE13} relations have an accuracy of 0.5 subtypes. We also compare the EFOSC spectrum to optical M dwarf standard spectra
from \cite{KIRKPATRICK91} and find a best visual match to types M1.0/M1.5.  We perform a similar comparison of our DBSP spectra to M dwarf standards and 
find a consistent best match to the M1 standard GJ 229 (Fig.~\ref{DBSP}). In the near-IR $K$-band, the H$_2$0-$K2$ index measures 
temperature sensitive water opacity and is calibrated for SpTy's M0-M9 \citep{ROJAS12}. We use our SpeX spectra to measure this index in K2-26 and 
K2-9 and find SpTy's of M1.0 and M2.5, respectively. The H$_2$0-K2/SpTy relation has a systematic scatter of 0.6 subtypes. Following these results,
we adopt a SpTy of M1.0 $\pm$ 0.5 for K2-26 and M2.5 $\pm$ 0.5 for K2-9. Our index based measurements are consistent with the visual best 
matches to M dwarf standards (e.g. Figures~\ref{SpeX_fig1} and~\ref{SpeX_fig2}) and are also consistent with SpTy estimates using the stars' optical and near-IR colors
 \citep{PECAUT13}\footnote{Throughout this work, we use the expanded table available on 
Eric Mamajek's webpage: \url{http://www.pas.rochester.edu/~emamajek/
EEM\_dwarf\_UBVIJHK\_colors\_Teff.txt}}. 

Following \citet{CROSSFIELD15}, we use our SpeX spectra to estimate the fundamental parameters of metallicity ($\mathrm{[Fe/H]}$), effective temperature ($\mathrm{T_{eff}}$),
radius (R$_*$), and mass (M$_*$) for K2-26 and K2-9 using the methods presented in \citet{MANN13_temp} and \cite{MANN13_metal}. 
In these works, metallicity is estimated using spectroscopic index and equivalent width based methods \citep{ROJAS12, TERRIEN12, MANN13_metal} that were 
calibrated using a sample of M dwarfs having wide, co-moving FGK companions with well determined [Fe/H]. We use IDL software made publicly available 
by A. Mann\footnote{\url{https://github.com/awmann/metal}} to calculate the metallicities of K2-26 and 
K2-9 in the $H$ and $K$ bands. We average the $H$ and $K$ 
metallicities and add the measurement and systematic uncertainties in quadrature to arrive at the final values. We find K2-26 has 
$\mathrm{[Fe/H] = -0.13 \pm 0.15}$ and K2-9 has $\mathrm{[Fe/H] = -0.25 \pm 0.20}$. Thus neither star is metal-rich.

Effective temperature, radius, and mass are calculated using temperature sensitive spectroscopic indices in the $JHK$-bands \citep{MANN13_temp} 
and empirical relations calibrated using nearby, bright M dwarfs with interferometrically measured radii \citep{BOYAJIAN12}. We calculate $\mathrm{T_{eff}}$ in 
the $JHK$-bands and average the results. Conservative $\mathrm{T_{eff}}$ uncertainties are estimated by adding in quadrature the rms scatter in the $JHK$-band values 
and the systematic error in the empirical fits for each band \citep{MANN13_temp}. The stellar radii, masses, luminosities, and densities are computed 
using publicly available software from A. Mann\footnote{\url{https://github.com/awmann/Teff_rad_mass_lum}}. The resulting fundamental parameters are listed in
Table~\ref{properties_table}. The larger relative uncertainties in R$_*$ and M$_*$ for K2-9 are a result of the poorer empirical fits in the 
\cite{MANN13_temp} relations due to having relatively few calibrators at low temperatures. {\color{black} \cite{VANDERBURG_15_cat} estimated $\mathrm{T_{eff}}$ and 
R$_*$ for K2-26 using its $V-K$ color and found results consistent with ours. They also estimated $\mathrm{T_{eff}}$ and 
R$_*$ for K2-9, this time using its $H-K$ color, and found values consistent with ours at $\sim$2$\sigma$.  Additionally, 
\cite{MONTET15} estimated the fundamental parameters of K2-9 using 
broadband photometry and model fits.  Our spectroscopic parameters are consistent with theirs in all cases within 1$\sigma$ uncertainties, but our 
nominal values of mass, radius, and metallicity are all systematically larger.} Photometric distances to the stars are estimated by calculating 
the distance moduli for their spectral types from the color-temperature conversion table of \citet{PECAUT13}. We estimate K2-26 
lies at $93 \pm 7$ pc and K2-9 at $110\pm12$ pc, just at the outer boundary of the extended solar neighborhood. 

\subsubsection{High-Resolution Spectroscopy}

We searched for tight, spectroscopic binary companions or background stars at very close 
angular separations in our HIRES spectra using the methodology of \citet{KOLBL15}. The \citet{KOLBL15} algorithm uses a library of more than 
600 HIRES spectra of stars with a range of $\mathrm{T_{eff}}$'s, log($g$)'s, and metallcities
to model the spectrum of the target star as the sum of two library templates and search for secondary lines. For high SNR targets, this method can detect 
companions within $\sim$0\farcs8 of the primary, with as little as $\sim$1\% of primary's flux in the $V$-band, and $\Delta$RV $>$ 10 km s$^{-1}$. 
The algorithm also measures the barycentric corrected primary RV via comparison to a standard  solar spectrum.

Neither {\color{black} of our K2-26 spectra nor the spectrum of K2-9 exhibit evidence for a close in, spectroscopic companion}. Our analyses exclude 
tight companions as faint as 3\% of the primary flux in the approximate $V$-band with $\Delta$RV $>$ 10 km s$^{-1}$. The lower flux limit corresponds to 
companions with $\Delta V \approx 3.8$. Assuming circular orbits and using the photometric relations in \cite{PECAUT13}, our $\Delta V$ and $\Delta$RV 
limits allow us to rule out companions {\color{black} $\sim$M4.5 and earlier at $\lesssim$0.7 AU separations for K2-26 and companions $\sim$M5.0 
and earlier at $\lesssim$0.7 AU separations for K2-9. We measure RV = $95.34 \pm 0.15$ km s$^{-1}$ and RV = $95.33 \pm 0.15$ km s$^{-1}$ for K2-26 
in February and November 2015, respectively. We also measure RV = $-31.02 \pm 0.15$ km s$^{-1}$ for K2-9. Our two RV measurements for K2-26 are 
separated by 281 days and are consistent within the 150 m s$^{-1}$ measurement uncertainty. These RV measurements are also consistent with the multi-epoch 
measurements spanning 28 days in \cite{VANDERBURG_15_cat}. The consistency of these long term RV measurements allows us to rule out RV accelerations due 
to companions below the sensitivity of our initial secondary line search. If we adopt our 150 m s$^{-1}$ HIRES measurement uncertainty as the maximum possible 
acceleration and assume circular orbits, our multi-epoch measurements rule out stellar mass companions at separations that overlap the limits from our LBT AO imaging (see \S\ref{AO_sec}) and a range of gas-giant companions that includes $\gtrsim$2 $\mathrm{M_{Jup}}$ at $\lesssim$0.25 AU and $\gtrsim$5 $\mathrm{M_{Jup}}$ at $\lesssim$1.5 AU.}

\subsection{Activities, Kinematics, Ages, and Surface Gravities}

\begin{figure}[!htb]
\epsscale{1.0}
\plotone{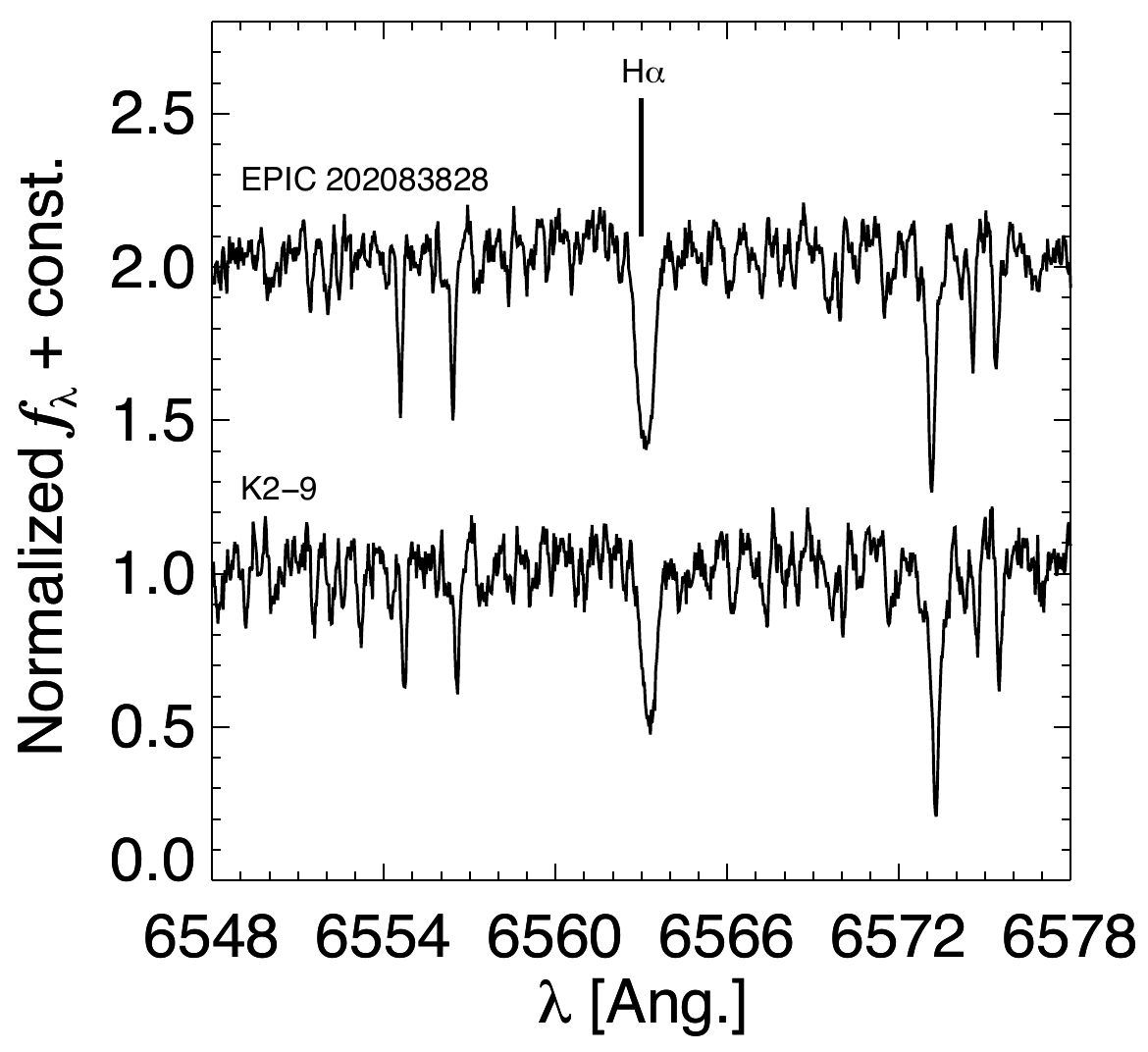}
\caption{RV corrected HIRES spectra of K2-26 (EPIC 202083828) and K2-9 centered on the H$\alpha$ line at 6563~\AA. The line is seen in absorption in both stars, indicating they are relatively inactive and likely $\gtrsim$1 Gyr old.\label{HIRES_fig}}
\end{figure}

The $\sim$6000 - 9000~\AA~(0.6 - 0.9 $\mu$m) region of M dwarf spectra provide access to several features sensitive to surface gravity and magnetic activity. Prior to the transition
to fully convective interiors ($\lesssim$M4), M dwarfs lose angular momentum to a steady stellar wind and their rotation rates decrease over time. 
This decrease in rotation rate leads to a loss in dynamo driven magnetic activity and a subsequent loss of high energy emission over time.  
Here we focus on emission from the 6563~\AA~H$\alpha$ line as an activity indicator \citep{WEST08, WEST11}
to place constraints on the ages of K2-26 and K2-9.  Our HIRES spectra of both K2-26 
and K2-9 provide access to the H$\alpha$ line at high resolution where it is seen in absorption in both stars (Figure~\ref{HIRES_fig}). We used the IDL software 
\texttt{line\_eqwidth}\footnote{\url{http://fuse.pha.jhu.edu/analysis/fuse\_idl\_tools.html}} to estimate equivalent widths (EWs) of $0.46 \pm 0.02$~\AA~and $0.34 \pm 0.01$~\AA~for K2-26 and K2-9, respectively. The H$\alpha$ EWs suggest that both stars are relatively inactive; consistent with field age early M dwarfs \citep{WEST08}. Our EFOSC2
and DBSP spectra of K2-26 cover the H$\alpha$ line where it is also seen in absorption.  
 
As an additional check of the stars' activity levels, we searched for excess ultraviolet (UV) emission using data from the NASA $GALEX$ satellite \citep{MARTIN05}. Like H$\alpha$, 
UV is tracer of magnetic activity in late-type stars and can be used to place limits on their ages \citep{SHKOLNIK11, STELZER13, JONES15}. We 
searched the $GALEX$ data using a 
10$^{\prime\prime}$ radius centered on our targets in the GalexView Web Tool\footnote{\url{galex.stsci.edu/GalexView/}}. No $GALEX$ observations are available for 
K2-26. K2-9 was observed by $GALEX$ during the 
Medium Imaging Survey \citep[MIS,][]{MARTIN05} in the far- and near-UV (FUV, NUV) bands. The star was detected at $\sim$4$\sigma$ and $\sim$5$\sigma$ in the
FUV and NUV, respectively. The MIS observations spanned 2006 October to 2009 April.
Following \citet{SHKOLNIK11} and \citet{SCHLIEDER12_NYMG}, we calculate the ratio of the 
FUV and NUV flux densities to the 2MASS $J$ and  $K_s$-band flux densities and compare to samples of M dwarfs with known ages. The flux density ratios indicate that 
K2-9 has FUV and NUV emission consistent with stars of similar SpTy having relatively
large excess emission and young ages. When compared to the M dwarf samples with known ages in \citet{SHKOLNIK14_HAZMAT}, the fractional UV excesses 
suggest that K2-9 is at most as old as the Hyades \citep[600-800 Myr,][]{PERRYMAN98, BRANDT15_a, BRANDT15_c}. It is possible that one of the epochs of the $GALEX$ 
measurements caught the star during a flare or other transient period of heightened activity. 
UV flare events were observed in $\sim$3\% of field M dwarfs in the variability survey of \citet{WELSH07} 
and \citet{FRANCE13} observed flares with timescales of 100-1000s in Hubble Space Telescope UV spectra of field age M dwarf planet hosts. These events are a likely 
contributor to the measured ranges of UV activity in M dwarf samples of known age \citep[up to 2 orders of magnitude,][]{SHKOLNIK14_HAZMAT}. Therefore, the evidence for 
at least transient strong UV emission from K2-9 is intriguing and may affect the properties of its planet, but when considered along with 
the large scatter of M dwarf UV excesses and the star's lack of H$\alpha$ emission, it does not indicate that the star is strikingly young. We note that the $\sim$80 day 
$K2$ light curve of K2-9 does not exhibit convincing evidence for magnetic spot modulated variability or strong flares. 

We investigate the kinematics of K2-26 and K2-9 by calculating their $UVW$ Galactic velocities following the methods outlined in 
\cite{JOHNSON87} updated to epoch J2000.0. We adopt a solar centric coordinate system where $U$ is positive toward
the Galactic center, $V$ is positive in the direction of solar motion around the Galaxy, and $W$ is positive toward the north Galactic pole. Using the measured proper motions and
RV's and estimated distances, we calculate $UVW_{K2-26}$ = (-91.7, -52.0, -28.7) $\pm$ (2.1, 3.6, 2.2) km s$^{-1}$ and $UVW_{K2-9}$ = (-86.3, -8.1, -41.6) $\pm$ 
(9.6, 3.1, 2.0) km s$^{-1}$. These estimates yield total Galactic velocities $S_{K2-26} = 109.3$ km s$^{-1}$ and $S_{K2-9} = 96.1$ km s$^{-1}$. Both stars have $S$ 
consistent with the statistically older, kinematically hotter, thick disk population following the kinematic sub-divisions of \cite{BENSBY10}. Their kinematics are thus 
consistent with other old M dwarfs.  
Alone, neither the lack of H$\alpha$ emission nor the large Galactic velocities of K2-26 and K2-9 place strong constraints on their ages. 
However, when combined, these observations suggest both stars are $\gtrsim$1 Gyr old. The observed UV excess of K2-9 warrants further consideration 
and is detailed in the context of its planet in \S~\ref{discussion}.

Spectroscopic akali lines are sensitive to electron pressure in stellar atmospheres in the sense that 
increases in pressure lead to lines with broader wings (pressure broadening), thus in 
low pressure (low gravity) atmospheres, alkali lines are comparatively weak \citep{SCHLIEDER12_grav}. To verify that K2-26 and K2-9 are dwarf stars with 
high surface gravities we investigate the gravity sensitive Na I doublet near 8190~\AA. After reducing our spectra to a resolution of R$\sim$900, we measured the Na I index as defined by \cite{LYO04} and compared 
 to samples with known surface gravities. For K2-26 we use the EFOSC2 spectrum to measure an Na I index of 1.064 $\pm$ 0.023. We prefer the EFOSC2 spectrum over the SpeX spectrum for this measurement since the SNR is $\sim$6$\times$ greater at 8200~\AA. For K2-9, we have only the SpeX spectrum with SNR$\sim$20 
at 8200~\AA~and measure an Na I index of 1.144 $\pm$ 0.129. The Na I uncertainties are estimated using MC methods. When compared to the dwarf, young, and giant
samples in \citet{LAWSON09}, both K2-26 and K2-9 are consistent with the field M dwarf sequence within uncertainties.  
This is reinforced by our initial target selection using reduced proper motion diagrams (which removes giants) and the visual matches of their SpeX spectra to M dwarf standards 
(Figs.~\ref{SpeX_fig1} and~\ref{SpeX_fig2}).

\subsection{Imaging Analyses}

\subsubsection{Adaptive Optics Imaging}\label{AO_sec}

In our LBT/LMIRcam imaging, K2-26 was measured with a resolution of 0\farcs116 (FWHM) and appears single at that limit. No other stars were detected 
in the 4\farcs0 region of full dither overlap. We estimate the sensitivity to faint companions and background stars by injecting fake point sources with SNR=5 into the 
final image at separations $N \times$ FWHM where $N$ is an integer. The 5$\sigma$ sensitivities as a function of separation are shown in the left panel of 
Fig.~\ref{AO_fig}. Our LMIRcam imaging is sensitive to stars with $K_s$-band contrast $\Delta K_s = 3.1$ mag at 0\farcs1 separation and $\Delta K_s = 7.2$ mag at 
$\ge0\farcs5$ separation.  At the distance of K2-26 and ages $\gtrsim$1 Gyr, these magnitude limits correspond to $\sim$M5 and earlier companions at 
$\gtrsim$9 AU and substellar companions at $\gtrsim$47 AU \citep{BARAFFE15, PECAUT13}.

Our Robo-AO images of K2-26 measure the star at the diffraction limit of 0\farcs15 \citep{BARANEC14}. We detect an additional star with 
$\Delta LP600 =  5.02 \pm 0.07$ mag at 5\farcs53 to the southeast. In the inset of the center panel in Fig.~\ref{AO_fig}, we show K2-26 with a hard stretch
and highlight this star with a red circle. This source falls within the 2 pixel (8$^{\prime\prime}$) software aperture used to extract the $K2$ photometry.  However,  
with $\Delta LP600 \sim$ 5 ($LP600 \approx Kep$), it contributes only $\sim$1\% of the flux in the aperture and would have to harbor a deeply eclipsing stellar companion to 
account for the measured $\sim$0.3\% transit depth. We discuss this possibility further in \S~\ref{validation_sec}. We place limits on the presence 
of other stars in the Robo-AO image by masking out the faint companion, replacing its flux with the median background at it's separation, and 
performing the multiple levels of automated companion search described in \citet{LAW14}. These searches identify no additional companion candidates and 
provide 5$\sigma$ sensitivity limits as a function of separation (center panel of Fig.~\ref{AO_fig}). Our Robo-AO imaging rules out stars with $\Delta LP600 \le 5$ mag
beyond the 2\farcs0 radius of our much more sensitive LMIRcam image. At the distance of K2-26, this sensitivity corresponds to $\sim$M5 companions at $\gtrsim$194 AU.  

The Keck/NIRC2 image of K2-9 reveals the star with a resolution of 0\farcs066 (FWHM) and it appears single at that limit. No other stars were detected 
in the full field of view of the image. We estimate the sensitivity to faint companions and background stars following the same procedure described for the LMIRcam
image of K2-26. The 5$\sigma$ sensitivities as a function of separation are shown in the right panel of 
Fig.~\ref{AO_fig}. Our NIRC2 imaging is sensitive to stars with $K_p$-band contrast $\Delta K_p = 3.5$ mag at 0\farcs1 separation and $\Delta K_p = 7.5$ mag at 
$\ge0\farcs5$ separation.  At the distance of K2-9 and ages $\gtrsim$1 Gyr, these magnitude limits correspond to $\sim$M7 companions at 
$\gtrsim$11 AU and substellar companions at $\gtrsim$55 AU \citep{BARAFFE15, PECAUT13}. {\color{black} Since both K2-26 and K2-9 are within approximately 100 pc of the Sun, reddening by interstellar dust is assumed to have a negligible effect and is not taken into account in our estimates of companion detection limits.}

\subsubsection{Archival DSS and SDSS Imaging}\label{archival_sec}

Both K2-26 and K2-9 have total proper motions $>$0\farcs1 yr$^{-1}$. Over the time baseline between their POSS I and SDSS observations 
K2-26 has moved $\sim$6\farcs3 and K2-9 has moved $\sim$9\farcs4. We display this large transverse motion in Fig.~\ref{archival_fig} 
where the sub panels are centered on the stars' epoch 2015 coordinates (magenta circles). The left column of Fig.~\ref{archival_fig} reveals that there are 
no background sources at the current positions of the stars down to the POSS I $B$ limit of 21.1 mag \citep{ABELL55}. {\color{black} We also find no background 
sources down to the POSS I $R$ limit of 20.0 mag \citep{ABELL66}. If there are background sources at the current
position of K2-26 they must be $B\sim5$ mag and $R\sim7$ mag fainter. For K2-9, undetected background sources must be $B\sim4.5$ mag and $R\sim6$ mag fainter.}
The archival images also reveal a star to the southeast of K2-26.  This source, 
SDSS J061649.67+243539.9, lies 6\farcs3 away in the SDSS epoch (2006.9) and has a relatively small proper motion of 32.9 mas yr$^{-1}$ to the north-northwest 
\citep{ABAZAJIAN09, ROESER10} and is thus not co-moving with K2-26. The position angle and separation in the SDSS image are consistent 
with this star being the same as that revealed in our Robo-AO image. 
SDSS J061649.67+243539.9 is 4.2 mags fainter than K2-26 in the SDSS $g$-band and its SDSS7 and 2MASS photometry are consistent with a 
K5 $\pm$ 2 spectral type \citep{PECAUT13, KRAUS07} assuming no reddening. With $J$=16.37, a $\sim$K5 star would be $\sim$1.8 kpc distant and lie far behind 
K2-26. The SDSS $g$ image of K2-9 reveals a very faint source at only 3\farcs5 to the northeast where the object appears as an extension to the 
K2-9 intensity distribution. This source, SDSS J114503.63+000021.4, has $r$ = 22.8 and is undetected in other archival surveys (DSS, 2MASS, etc.). 
Thus, no proper motion measurements are available. The source is also undetected in our NIRC2 images of K2-9 which have a 5$\sigma$ sensitivity at wide 
separations of $K \sim 19$ mag. Regardless of the nature of this object, we estimate that it contributes negligible flux ($\sim$0.1\%) to the $K2$ software aperture.

\subsection{Light Curve Validation and False Positive Probabilities}\label{validation_sec}
 
After \texttt{TERRA} identifies candidate transits, it runs a suite of diagnostics to vet out possible astrophysical false positives such as eclipsing binaries,
spot modulation, or periodic stellar variability. K2-26 and K2-9 both passed these tests and were subject to further extensive testing to explore 
centroid motions in and out of transit, difference imaging analyses, and pixel correlation images \citep[e.g.] []{BRYSON13}. Both stars pass these further checks and we find it 
unlikely that the observed transits associated with K2-26 occur around the faint star observed to the southeast in the Robo-AO and 
archival images. This is corroborated by independent $Spitzer$ detections of the transit using a software aperture 
that excludes the nearby faint star (see Beichman et al.~2015, submitted). 

Other possible scenarios that could give rise to the observed transits are an unblended EB, 
unresolved bound companions hosting their own transiting stellar or planetary companion, or background EBs. 
{\color{black} Our Keck/HIRES and AO imaging analyses have ruled out a wide range of bound companions.} Given these constraints, the radius of a potential planet around any 
undetected bound companion would have to be 
improbably large to produce the measured transit depths when considering the dilutions from K2-26 and K2-9 \citep{DRESSING13, DRESSING15}. 
The lack of detectable stars at the current positions of K2-26 and K2-9 in POSS I images also places strong constraints on unresolved background 
eclipsing binaries not ruled out by our spectroscopy and imaging. 

As a final check, we estimate the likelihood that our observed transits are false positives rather than {\em bona fide} 
planetary systems using the open source false positive probability (FPP) calculator \texttt{vespa} 
\citep{MORTON12, MORTON15_vespa}\footnote{\url{http://github.com/timothydmorton/vespa}}. To calculate FPPs, this software package compares the typical 
light curve shapes of a distribution of astrophysical false positive scenarios to the observed transit light curve and combines 
that information with prior assumptions about stellar populations, multiplicity frequencies, and planet occurrence rates. The false positive scenarios 
tested are: an unblended EB, a blended background EB, a hierarchical companion EB, and the `double-period' EB scenario which is newly 
implemented in \texttt{vespa}.  {\color{black} A key input to \texttt{vespa} in the standard EB scenarios is the secondary depth constraint, $\delta_{sec, max}$,
which is the deepest secondary event allowed at all phases. This is determined by masking the transit signal of the candidate planet and searching the light 
curve for the most significant signal at the same period \citep[see][]{MORTON12, MONTET15}.} The newly implemented `double-period' scenario in \texttt{vespa}
is the hypothesis that the transit signal is caused by an EB (either unblended, background, or hierarchical) 
at double the measured period where the primary and secondary eclipses are the same depth. This case cannot be subject to the same secondary depth constraint
as the others; rather, it is subject to an odd-even constraint that requires the primary and secondary 
eclipses have depths within $\sim$3$\sigma$ of the photometric uncertainty in the phase-folded light curve.

As additional inputs to \texttt{vespa}, we used our phase folded light curves, the stellar photometry from APASS, 2MASS, and WISE, the stars' physical parameters listed in 
Table~\ref{properties_table}, our near-IR AO contrast curves, {\color{black} the constraints on background stars from the POSS I $R$-band archival images,} and our HIRES RV 
constraints implemented as a velocity contrast curve in \texttt{vespa} \citep[e.g.][]{MARCY14}. {\color{black} We ran \texttt{vespa} within 
5\farcs0 and 12\farcs0 of K2-26 and K2-9, respectively. The K2-26 aperture was chosen to exclude the faint companion from our Robo-AO image 
which we have already ruled out as the transit host. In the absence of close companions, the K2-9 aperture was chosen to be consistent with 
the $K2$ photometric aperture.} Table~\ref{FPPs} lists the FPPs from each of the 
tested false positive scenarios. {\color{black} We find total FPPs of $3.2 \times 10^{-3}$ and $2.4 \times 10^{-5}$ for K2-26 and K2-9 respectively.}  
The estimated FPPs are sufficiently low that we consider both systems to be validated exoplanets. Our \texttt{vespa} input files are available upon request. 
 
\begin{table}
\begin{center}
\caption{False Positive Probability Calculation Results  \label{FPPs}}
\begin{tabular}{lcc}
\hline\hline
Param  &      K2-26b & K2-9b \\
\hline
   $\delta_{sec, max}$ [ppt]$^{a}$ &  0.24 & 0.47 \\
       Pr$_{EB}$ &   $<10^{-4}$               &  $<10^{-4}$         \\
       Pr$_{DPEB}$ &    $<10^{-4}$                                 & $<10^{-4}$         \\
 Pr$_{BEB}$&   $2.4 \times 10^{-3}$                           & $<10^{-4}$                   \\
  Pr$_{DPBEB}$ &   $<10^{-4}$                            & $<10^{-4}$          \\
  Pr$_{HEB}$ &   $<10^{-4}$                           & $<10^{-4}$         \\
   Pr$_{DPHEB}$&   $<10^{-4}$             & $<10^{-4}$  \\
       $f_p$$^{b}$ &    0.21                         & 0.20           \\
  \hline
       $FPP$ &    $3.2 \times 10^{-3}$                   & $2.4 \times 10^{-5}$      \\
  \hline
Planet?    &    Yes      &    Yes \\
\hline
\label{FPPs}
\end{tabular}
\tablecomments{$^{a}$maximum secondary depth, $^{b}$integrated planet occurence rate}
\end{center}
\end{table}

 \subsection{Planet Parameters}

{\color{black} We analyze the time-series photometry for these systems using an approach similar to the one described by \cite{CROSSFIELD15} which relies on the
\texttt{emcee} Markov Chain Monte-Carlo (MCMC) package \citep{MACKEY13}.  
For these planets, we estimated their parameters using two different light curve analysis packages: JKTEBOP
\citep{SOUTHWORTH04, SOUTHWORTH11} and BATMAN \citep{KREIDBERG15}.
We also used both linear and quadratic limb-darkening (LD) relations. For the linear LD relation, we impose Gaussian priors on
the LD parameter, determined by examining all linear LD terms tabulated by \citet{CLARET12} that satisfy $3300\le T_{eff}
\le3700$~K and $\log_{10} g \ge 4.5$. For the prior, we take the mean and twice the standard deviation of these values, in order to account
for possible systematic uncertainties in the models \citep{ESPINOZA15}. For the quadratic LD relation, we estimated parameters using the LDTk package \citep{PARVIAINEN15, PARVIAINEN15_LDTK}\footnote{{\url{https://github.com/hpparvi/ldtk}}}. LDTk calculates custom LD profiles and coefficients using a library of PHOENIX spectra. The uncertainties on the LD parameters are propagated from the uncertainties on the stellar parameters.   For each light curve analysis package and both linear and quadratic LD, we assumed circular orbits and calculated the fit parameters. In every case, we find consistent results. Thus, we adopt the fit parameters from JKTEBOP using linear limb-darkening for the remainder of our analyses (Table~\ref{tab:planet}).}

These fits provide planetary radii R$_{K2-26b}$ = $2.67^{+0.46}_{-0.42}$ $R_{\oplus}$ and R$_{K2-9b}$ = 
$2.25^{+0.53}_{-0.96}$ $R_{\oplus}$. Thus, both planets are small and have estimated radii near the transition
between Earth-like rocky bodies and Neptune-like bodies with large gaseous envelopes \citep{MARCY14, WEISS14, ROGERS15}. Our transit fits also provide
an estimate of the stellar density, $\rho_{*}$. For K2-26, the stellar density from the fit is inconsistent with the density inferred from our spectroscopic
constraints on the star's mass and radius (Table~\ref{tab:planet}). This result is driven by the measured transit duration of 4.73 h; $\sim$1.8 h longer 
than expected for a circular orbit with a low impact parameter. The long transit duration can be explained either by the planet orbiting a low density (giant) star 
with a large radius or by the planet having an eccentric orbit where the transit is observed away from periapse. This phenomenon is known 
as the photoeccentric effect \citep{DAWSON12}. Since our spectroscopic and other analyses conclusively demonstrate that K2-26 is a dwarf, not a giant, planetary 
eccentricity likely causes the long transit duration. {\color{black} In the case of K2-9, the larger uncertainties on the stellar parameters result in a larger 
uncertainty on the stellar density in the fit. Although the nominal value is inconsistent with the spectroscopic constraints, the full uncertainty range is 
consistent. We also find that the measured transit duration is consistent with expectations for an approximately circular orbit.}

To further test these hypotheses, we examine the photoeccentric effect in post-processing of the
posterior distributions to the MCMC transit analysis that assumed a circular orbit. The final unimodal distribution hints 
that K2-26 may have an eccentric orbit with a lower eccentricity 
limit $e>0.14$ with 95\% confidence ($e>0.01$ at 97\% confidence). {\color{black} In addition, the posterior on the argument of periapsis, $\omega$, peaks 
near 270$^{\circ}$, consistent with our qualitative assessment that the transit occurs away from periapse. The same analysis of K2-9b provides a lower eccentricity 
limit of $e>0.05$ at 95\% confidence. This constraint supports our previous assessment that the orbit of K2-9b is approximately circular. We 
also performed a photoeccentric analysis on several dozen other planet candidates from our \emph{K2} sample and found only two other systems with evidence for eccentricity.
In contrast to K2-26b, these candidates exhibited transit durations shorter than expected for a circular orbit with $\omega$ close to 90$^{\circ}$.} As a final note,
we mention that any photoeccentric analysis is subject to measurement bias such that any constraints on eccentricity are always positive and non-zero. 
Figs.~\ref{3828_lightcurve_fig} and~\ref{5501_lightcurve_fig} show the resulting photometry and best-fit models and 
Table~\ref{tab:planet} summarizes the final planetary parameters and uncertainties.

\begin{table}
\begin{center}
\caption{Summary of Planet Properties  \label{tab1}}
\begin{tabular}{llcc}
\hline\hline
Param  & units & K2-26b & K2-9b \\
\hline
   $T_{0}$ & $MBJD_{TDB}^{a}$  & $1942.1659^{+0.0028}_{-0.0021}$ & $1989.6712^{+0.0025}_{-0.0033}$ \\
       $P$ &          d            & $14.5665^{+0.0016}_{-0.0020}$               & $18.4498^{+0.0015}_{-0.0015}$         \\
       $i$ &        deg            & $88.4^{+1.2}_{-1.5}$                                & $87.983^{+1.593}_{-0.080}$         \\
 $R_P/R_*$ &         \%      & $4.71^{+0.37}_{-0.22}$                           & $6.669^{+0.060}_{-1.307}$                    \\
  $T_{14}$ &         hr         & $4.73^{+0.25}_{-0.12}$                            & $2.8397^{+0.0074}_{-0.5492}$          \\
  $T_{23}$ &         hr         & $4.11^{+0.15}_{-0.44}$                           & $0.62^{+1.38}_{-0.27}$         \\
   $R_*/a$ &         --           & $0.0483^{+0.0186}_{-0.0076}$             & $0.0380^{+0.0011}_{-0.0208}$   \\
       $b$ &         --              & $0.57^{+0.24}_{-0.39}$                         & $0.9261^{+0.0080}_{-0.4929}$           \\
       $u$ &         --              & $0.566^{+0.048}_{-0.047}$                   & $0.579^{+0.105}_{-0.070}$      \\
  $\rho_{*, circ}$$^{b}$ & g~cm$^{-3}$       & $0.79^{+0.53}_{-0.49}$   & $1.010^{+9.870}_{-0.096}$         \\
   $\rho_{*, spec}$$^{c}$ & g~cm$^{-3}$     & $3.92^{+1.43}_{-1.43}$              & $9.88^{+4.25}_{-4.25}$        \\
       $a$ &         AU            & $0.0962^{+0.0054}_{-0.0061}$               & $0.091^{+0.013}_{-0.016}$      \\
     $R_P$ &      $R_{\oplus}$      & $2.67^{+0.46}_{-0.42}$                          & $2.25^{+0.53}_{-0.96}$                  \\
 $S_{inc}$ &      $S_{\oplus}$      & $\sim$5.8                                                  & $1.36^{+1.59}_{-0.81}$             \\
$T_{eq}$ & K                     & $\sim$430						& $314^{+67}_{-64}$                   \\
     $R_*$ &  $R_\odot$      & $0.520^{+0.080}_{-0.080}$                    & $0.31^{+0.11}_{-0.11}$       \\
     $M_*$ &  $M_\odot$     & $0.560^{+0.100}_{-0.100}$                     & $0.30^{+0.14}_{-0.13}$         \\
\hline
\label{tab:planet}
\end{tabular}
\tablecomments{$^a$BJD - 2454833; $^b$Stellar density from transit fits assuming circular orbits; $^c$Stellar density from spectroscopic stellar parameters.}
\end{center}
\end{table}

\section{Discussion}\label{discussion}

\subsection{K2-26b: A temperate sub-Neptune with evidence for eccentricity}

K2-26b was the only candidate M dwarf planet found in our search of the $K2$ C0 data and is the first 
validated planet from that field. $K2$ C0 was anticipated to be the first full length campaign for the spacecraft in 
its new observing mode, but fine guiding control was only achieved for $\sim$35 days thereby limiting useful data to this period \citep{VANDERBURG14_C0}. 
The discovery of a small planet with a 14.57 day orbital period in only 35 days of data is both fortuitous and a 
testament to the quality of the $K2$ data.  K2-26b bleongs to a class of planets known as sub-Neptunes: planets smaller than Neptune with 
substantial H/He atmospheres \citep{MARCY14, WEISS14, ROGERS15}. We estimate 
the mass of the planet and the likelihood {\color{black} that it is more dense than 100\% silicate rock} following the probabilistic approach described in \cite{WOLFGANG15_2} and \cite{WOLFGANG15_1}.
This approach uses a sample of known, small planets with measured masses and radii, interior structure models, and hierarchical Bayesian modeling to fit a mass-radius  
relation to the data that includes measurement and systematic errors. 
Using software provided by A. Wolfgang\footnote{\url{https://github.com/dawolfgang/MRrelation}}, we find M$_{K2-26b}$ = 
9.4 $\pm$ 3.3 M$_{\oplus}$ and $<$3\% {\color{black} probability that the planet is more dense than silicate rock}. The deterministic fit described in \cite{WEISS14} provides 
a nominally smaller, but still consistent, mass of $\sim$6.7 M$_{\oplus}$. 

\begin{figure}[!htb]
\epsscale{1.0}
\plotone{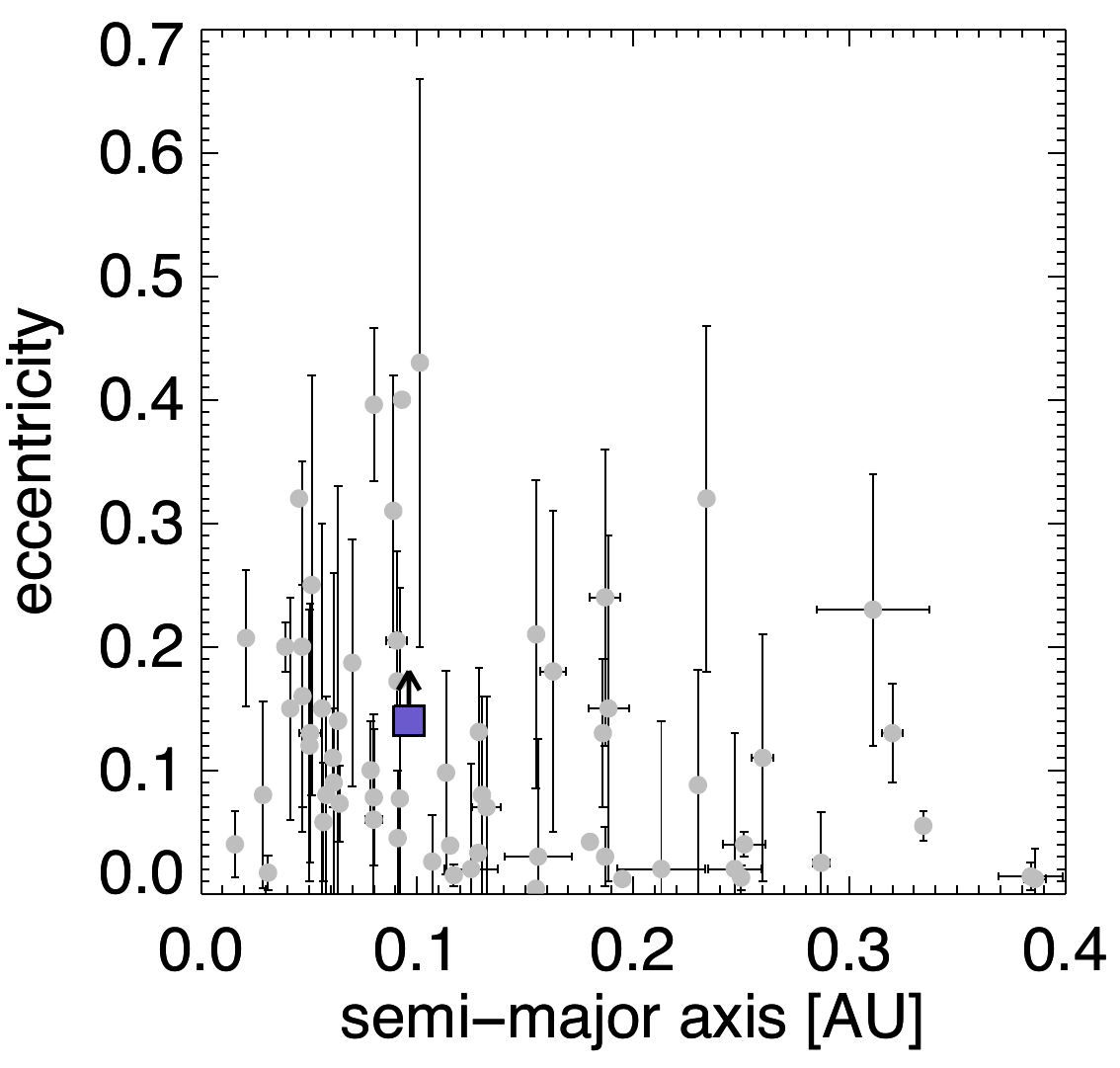}
\caption{Eccentricity vs. semi-major axis for planets in the NASA Exoplanet Archive that have separations less than 0.4 AU, are less massive than Neptune, 
and have non-zero eccentricities. Our K2-26b 95\% confidence lower eccentricity limit is shown as a slate blue square. \label{eccen_fig}}
\end{figure}

To further investigate the hint of eccentricity in K2-26b, we searched the NASA Exoplanet 
Archive\footnote{\url{http://exoplanetarchive.ipac.caltech.edu/}} \citep{AKESON13} 
for similar small planets to compare its eccentricity constraints to known 
systems. The database contains 63 transit or RV detected planets with $M_{p} < 1~M_{Nep}$, $a < 0.4$ AU, and $e > 0$. In Figure~\ref{eccen_fig}, 
we show the eccentricities vs.~semi-major axes of these planets along with our K2-26b 95\% confidence lower limit. 
Our K2-26b limit is larger than 65\% of the sample. Only about 20\% of the sample has eccentricities $\gtrsim0.2$. These data
are consistent with previous studies that found close in exoplanets have average eccentricities
smaller than planets on wider orbits; with most having $e < 0.2$ \citep{JACKSON08_tidal}. This is likely due to tidal dissipation of both the eccentricity and 
semi-major axis over time which circularizes the orbits \citep{RASIO96}. Thus, since K2-26b may have $e > 0.14$, 
the evolution of its orbit and its internal heat may be affected through tidal dissipation \citep{JACKSON08_heating}. 
 
In the process of tidal dissipation, both the eccentricity and semi-major axis of a planet are reduced as orbital energy is converted into internal heat
via tidal stresses. Thus, the rates of change of eccentricity, semi-major axis, and tidal heating are strongly coupled and only instantaneous values 
can be calculated outside of time dependent numerical integrations. We use the relations provided in \cite{JACKSON08_heating, JACKSON08_tidal} 
to investigate these effects where we assume that the eccentricity is equal to our inferred 95\% lower limit, $e = 0.14$. In this scenario, we estimate the
eccentricity and semi-major axis of K2-26b decrease by at least $\sim$3 $\times$ 10$^{-11}$ yr$^{-1}$ and $\sim$8 $\times$ 10$^{-13}$ 
AU yr$^{-1}$, respectively. These estimates correspond to a minimum tidal heating of $\sim$600 W m$^{-2}$. We also estimate the minimum orbit 
averaged incident flux following \cite{BARNES08} and find $\mathrm{S_{inc} \gtrsim 5.4~S_{\oplus}}$. Thus, the estimated tidal heating contributes at least an additional 
$\sim$8\% of the incident flux resulting in a total flux of $\mathrm{S_{tot} \gtrsim 5.8~S_{\oplus}}$. This yields a zero albedo equilibrium temperature for 
K2-26b of $\mathrm{T_{eq} \gtrsim 430 K}$. The equilibrium temperature remains $\lesssim$500 K for eccentricities $\lesssim$0.8. Following the habitable 
zone (HZ) description of \cite{KOPPARAPU14, KOPPARAPU13}, the planet's current separation and our tidal dissipation timescale estimates 
indicate that K2-26b was likely never in the HZ of its host star. This holds true even for eccentricities much larger than $e = 0.14$. 

A plausible explanation for the non-zero lower limit on the planet's eccentricity is the influence of an undetected perturber. 
{\color{black} Our multi-epoch HIRES RV measurements rule out a portion of mass/semi-major axis parameter space for $>$1$~\mathrm{M_{Jup}}$ companions. 
However, additional bodies may still exist in the system and induce transit timing variations \cite[TTVs,][]{HOLMAN05, AGOL05} of K2-26b.} 
We searched for TTVs in our 3 $K2$ transits of the planet by fitting for the central time in each transit individually, then fitting a straight line to the transit times 
as a function of transit number. Significant TTVs would be apparent when comparing the 
observed time of each transit to the best-fit linear ephemeris. We see no evidence of significant TTVs in the $K2$ data of K2-26b.
However, if the transits times are slowly varying, TTV signatures may be undetectable over the short time baseline of the $K2$ C0 data ($\sim$35 days). 
Thus, future follow-up over a longer baseline is warranted and could still reveal evidence of a perturber via TTVs. Additional high-cadence photometry 
over a period longer than $\sim$35 days and high precision RV monitoring would also be useful to search for additional planets and for placing better constraints on the 
possible eccentricity of K2-26b. 

\subsection{K2-9b: A transition radius planet receiving Earth-like insolation orbiting a UV active star}

The full range of estimated radii of K2-9b are consistent with the planet straddling the transition region between planets with Earth-like, rocky/iron compositions
and planets with Neptune-like, volatile rich compositions \citep{ROGERS15}. We use the same methods from \cite{WOLFGANG15_2} to estimate its mass and 
probability of rocky composition. From our estimated radius and an average symmetric radius uncertainty, we find M$_{K2-9b}$ = 7.6 $\pm$ 4.1 M$_{\oplus}$ 
and $\sim$21\% probability that is has an Earth-like, rock/iron composition. The smaller, but still consistent, radius estimated by \cite{MONTET15} yields a 
mass of 4.7 $\pm$ 2.9 M$_{\oplus}$ and $\sim$52\% {\color{black} probability of being more dense than silicate rock}. Thus, K2-9b, like K2-3cd and K2-21b, is a transition radius planet 
having $\sim$equal likelihood of either composition given current constraints on the densities and model inferred compositions of small planets. 
The deterministic mass-radius relation of \cite{WEISS14} provides a mass of $\sim$4.9 M$_{\oplus}$ using the average radius estimated from our 
analyses and \cite{MONTET15}.

The estimated flux from its host star incident on any atmosphere harbored by K2-9b is consistent with the flux received by the Earth from the Sun, 
$\mathrm{S_{inc} = 1.36^{+1.59}_{-0.81}~S_{\oplus}}$. Its estimated semi-major axis places it just within the inner edge of the star's optimistic 
HZ \citep{KOPPARAPU14, KOPPARAPU13}. The incident flux corresponds to an equilibrium temperature $\mathrm{T_{eq} =  315^{+67}_{-64} K}$. 
In the case that the planet is rocky, and has favorable atmospheric, cloud, and surface
properties, liquid water could exist on K2-9b. This potential is particularly interesting when the measured UV flux of the host star is considered.  

Whether the large $GALEX$ UV flux of K2-9 is constant or transient, this emission dominates the photochemistry of any atmosphere its planet harbors. UV photons 
photodissociate key molecules including H$_2$O, CH$_4$, and CO$_2$ \citep{KASTING93, SEGURA10} and can lead to the formation of high level hazes. Such
hazes have been inferred from the transmission spectrum of GJ 1214b, a small planet orbiting an M dwarf \citep{CHARBONNEAU09, BEAN11, KREIDBERG14}. UV flux can also 
cause atmospheric loss via photoevaporation \citep[][and references therein]{LUGER15}. For the case of intransient UV emission, 
we estimate the flux from the strongest stellar UV emission line, 
the resonance line of hydrogen at 1215.7~\AA, Lyman-$\alpha$ (Ly$\alpha$). We calculate the Ly$\alpha$ flux (F$_{Ly\alpha}$) from K2-9's $GALEX$ 
UV fluxes using the relations in \cite{SHKOLNIK14_Lyman}. These fits were calibrated using reconstructed intrinsic Ly$\alpha$ fluxes from a sample of K and M dwarfs with 
UV spectra from the Hubble Space Telescope \citep[MUSCLES,][]{LINSKY13, FRANCE13}. All fluxes in these calculations are scaled to the 
surface of the star using our photometric distance and the measured stellar radius. We estimate the F$_{Ly\alpha}$ at the surface of K2-9 is 
$\sim$2.5 $\times$ 10$^6$ erg cm$^{-2}$ s$^{-1}$. This corresponds to a S$_{Ly\alpha}$ at the planet of $\sim$8500 erg cm$^{-2}$ s$^{-1}$;  
nearly 5$\times$ larger than the estimated S$_{Ly\alpha}$ at the sub-Neptune GJ436b \citep{MIGUEL15, FRANCE13}.  If K2-9b harbors an atmosphere, the 
upper portions could therefore receive a UV flux where models predict significant chemical changes \citep{MIGUEL15} and possible
atmospheric loss \citep{LUGER15}. If the strong $GALEX$ UV flux was the result of a flare, the presumed atmosphere of the planet may be less affected \citep{SEGURA10}.

\subsection{Prospects for Follow-up}

Both K2-26b and K2-9b are small, temperate planets that orbit close to low-mass stars. We assume circular orbits
and adopt the nominal masses from the \cite{WOLFGANG15_1} probabalistic mass-radius relation to estimate
stellar RV semi-amplitudes of K$_{K2-26b}$ $\sim$ 3.6 m s$^{-1}$ and K$_{K2-9b}$ $\sim$ 4.1 m s$^{-1}$. These predicted reflex velocities are within 
the reasonable limits of current ground-based, high-precision spectrometers (e.g.~HIRES and HARPS), but the stars are too faint at 
visible wavelengths. {\color{black} These planets, along with others being discovered by \emph{K2} \cite[i.e.][]{HIRANO15}, may be ideal targets for next generation 
spectrometers operating in the IR where M dwarfs are brighter
 \citep[e.g.~CARMENES, SPIRou, IRD, HPF,][]{QUIRRENBACH14, ARTIGAU14, KOTANI14, MAHADEVAN14}}. High-precision RVs will
provide planet masses, filling in the critical transition region for temperate planets in the mass-radius-temperature diagram, provide additional constraints
on their orbital parameters, and also allow the detection of additional, non-transiting companions. We are pursuing a follow-up program with \emph{Spitzer} to refine 
orbit ephemerides and search for TTVs in small $K2$ planets transiting M dwarfs. {\color{black} The transit of K2-26b has been independently detected in 
this program and the \emph{Spitzer} data provides consistent transit parameters (Beichman et al.~2015, submitted).} Atmospheric characterization via transmission or emission spectroscopy is infeasible in the near future; both stars are too faint for transit spectroscopy to be practical with the \emph{JWST} 
\citep{COWAN15, BEICHMAN14}. However, such observations may be feasible with next generation, ground-based, 30m class telescopes \citep[][]{COWAN15}. 

\section{Conclusion}

We report on two small, temperate planets orbiting relatively nearby, cool stars observed during the $K2$ mission. Our detailed characterizations using our own observations
and other available data reveal K2-26b may have an eccentric orbit. We find that K2-9b lies in the optimistic HZ of its host star, but may receive 
high levels of UV flux that would likely affect the chemistry of its presumed upper atmosphere. Future observations can provide masses for the planets and critical constraints
on the transition between rocky and volatile rich bodies. The \emph{Kepler} spacecraft continues its legacy of discovery and is expected to observe many more
fields around the ecliptic during its $K2$ mission to reveal more small planets orbiting low-mass stars.  
\ \\
\acknowledgments

We thank the referee for their prompt, constructive report that has improved the quality 
of this manuscript. We thank the LBTI/LMIRcam instrument team for providing support during LBT 
observations. J.E.S thanks Tom Greene and Mike Werner for helpful discussions. The research of J.E.S was supported by an appointment to 
the NASA Postdoctoral Program at NASA Ames Research Center, administered by Oak 
Ridge Associated Universities through a contract with NASA. Support for E.A.P and  A.J.S. was provided 
by the National Aeronautics and Space Administration through Hubble Fellowship grant 
HST-HF2-51349 awarded by the Space Telescope Science Institute, which is operated by 
the Association of Universities for Research in Astronomy, Inc., for NASA, under contract NAS 5-26555.   
A.W.H. acknowledges NASA grant No.~NNX12AJ23G and S.L acknowledges NSF grant No.~AST 09-08419. 
C.A.B is grateful to Davy Kirkpatrick for his assistance
with planning and reduction of the Palomar Double Spectrograph
observations. The Large Binocular Telescope Interferometer is funded by NASA as part of its Exoplanet Exploration program. 
LMIRcam is funded by the National Science Foundation through grant NSF AST-0705296. The Robo-AO system was developed by
collaborating partner institutions, the California Institute of Technology, and the Inter-University Centre for Astronomy 
and Astrophysics, and supported by the National Science Foundation under Grant Nos.~AST-0906060, AST-0960343, and AST-1207891, the 
Mt.~Cuba Astronomical Foundation, and by a gift from Samuel Oschin. C.B. acknowledges support from the Alfred P. Sloan Foundation.
This work made use of the SIMBAD database
(operated at CDS, Strasbourg, France); NASA's Astrophysics Data System Bibliographic
Services; the NASA Exoplanet Archive and Infrared Science Archive, data products from the Two Micron All Sky Survey
(2MASS), the APASS database, the SDSS-III project, the Digitized Sky Survey, and the Wide-Field Infrared Survey Explorer. 
The authors wish to recognize and acknowledge the very significant cultural role and reverence that the summit of Maunakea 
has always had within the indigenous Hawaiian community. We are most fortunate to have the opportunity to conduct observations 
from this mountain.



{\it Facilities:} \facility{{\it Kepler}, K2, IRTF(SpeX), NTT(EFOSC2), LBT(LBTI/LMIRcam), Keck: I(HIRES), Keck: II(NIRC2), PO:1.5m (Robo-AO), PO:5.0m (Double Spectrograph)}



\bibliography{K2_refs}

\end{document}